\documentclass[man,floatsintext]{apa7}
\usepackage{graphicx}    
\usepackage{subcaption,pdfpages}
\usepackage{lipsum,amsmath,amsfonts,multirow}
\usepackage[american]{babel}
\usepackage{csquotes,enumitem}
\usepackage[style=apa,sortcites=true,sorting=nyt,backend=biber]{biblatex}
\DeclareLanguageMapping{american}{american-apa}
\hypersetup{
  colorlinks=true,
  linkcolor=blue,
  citecolor=blue,
  urlcolor=blue
}

\title{From Cognitive Relief to Affective Engagement: An Empirical Comparison of AI Chatbots and Instructional Scaffolding in Physics Education}
\shorttitle{From Cognitive Relief to Affective Engagement}

\authorsnames[{1,$\dagger$},{1,$\dagger$},{1,$\dagger$},{2,3},1]{E. Becker, J. Wünsche, J. M. Veith, J. Schrader, P. Bitzenbauer}

\authorsaffiliations{{Institute of Physics Education, Leipzig University, Leipzig, Germany},
{L3S Research Center, Leibniz University Hannover, Hannover, Germany},
{CAIMed – Lower Saxony Center for Artificial Intelligence and Causal Methods in Medicine, Hannover, Germany}\\
$^\dagger$ These authors contributed equally}

\abstract{Providing effective, personalized support is critical for helping students overcome conceptual difficulties in physics. However, established scaffolding methods, such as structured tiered support, are often too resource-intensive for widespread implementation. Therefore, this study, investigates whether an easily adaptable, custom-configured AI chatbot can offer comparable affective benefits and cognitive relief. We conducted a quasi-experimental field study with 273 ninth-grade students in Germany. Classes were randomly assigned to solve a buoyancy problem using one of three conditions: an AI chatbot, a tiered support system, or traditional textbook-style explanations. We measured intrinsic and extraneous cognitive load and affective outcomes (enjoyment, hope, hopelessness, self-efficacy, situational interest) via research-validated questionnaires. Results revealed that both interactive support systems -- the custom-configured AI chatbot and tiered hints -- were significantly more effective than the textual support in reducing students' intrinsic and extraneous cognitive load. Furthermore, the AI chatbot yielded the most comprehensive affective benefits, demonstrating significant improvements across all measured affective dimensions, when compared to the textual support. While the chatbot consistently trended more positively than the tiered hints on affective measures, these differences were not statistically significant. These findings suggest that while structured guidance is key to managing cognitive load, the interactive and social nature of AI chatbots holds unique potential for simultaneously fostering positive affective experiences, marking a promising direction for developing effective and holistic learning support tools in physics education.}

\keywords{physics education, AI chatbot, cognitive load, motivation, achievement emotions, affective outcomes, instructional scaffolding}

\authornote{%
   \addORCIDlink{J. M. Veith}{0000-0002-2147-0349}\\
   \addORCIDlink{E. Becker}{0009-0009-0985-6335}\\
   \addORCIDlink{J. W\"{u}nsche}{0009-0007-4380-3590}\\
   \addORCIDlink{J. Schrader}{0000-0001-9473-4635}\\
   \addORCIDlink{P. Bitzenbauer}{0000-0001-5493-291X}

  Correspondence concerning this article should be addressed to J. M. Veith.\\  E-mail: joaquin.veith@uni-leipzig.de}

\begin{document}
\maketitle

\section{Introduction} 

Physics is widely recognized as a conceptually challenging subject. A primary difficulty for students is translating theoretical principles into concrete conceptual models, a skill that typically requires deliberate pedagogical support~\parencite{Maries2017}. Without such guidance, many novices resort to \textit{formula-based problem solving} rather than developing a deep understanding of the underlying physics~\parencite{Maries2017,Belland2017}.

To address this challenge, Differentiated Instruction (DI) is a well-established approach designed to tailor teaching to the diverse needs of all learners~\parencite{UNESCO2020,Tomlinson2014}. A pedagogical strategy for achieving DI is employing tiered activities~\parencite{Tomlinson2017}, a form of instructional scaffolding, which provides temporary, adaptive support to help a student succeed in a task that would otherwise be out of reach~\parencite{Belland2017,Wood1976}. In physics education, ``tiered support'' (in German: gestufte Lernhilfen) exemplifies this by providing students with a sequence of optional, increasingly explicit learning aids, allowing them to individually adapt the level of assistance they receive~\parencite{StaeudelWodzinski2010}. The effectiveness of such scaffolding approaches is well-documented, positively impacting not only academic achievement but also student motivation~\parencite{FrankeBraun2008,Haenze2007,SchmidtWeigand2008}. However, widespread application poses a significant practical challenge. While teachers recognize student heterogeneity, they are often constrained by high workloads and limited time~\parencite{Bosch2024}. This makes the extensive effort required to develop high-quality tiered materials for every task a prohibitive undertaking.

Generative AI chatbots, especially when pedagogically tailored through specific configurations, present a promising technological solution to this pedagogical dilemma. They have the potential to act as personalized tutors, capable of reducing teacher workload while offering scalable, on-demand support to students engaged in parallel, individualized learning~\parencite[cf.][]{Neumann2024, Liang2023}. In fact, current research increasingly demonstrates the potential of AI chatbots to enhance both cognitive and motivational outcomes in STEM education, acting as intelligent tutors or providing personalized scaffolding~\parencite{Wang2025,ChenChang2024,Ji2023,Ng2024}. Other studies have explored the application of generative AI in open-ended tasks such as programming, creating teaching aids, or design thinking~\parencite{Johnson2024,Huang2024,Ji2023,Yilmaz2023}.

However, it remains an open question how affective outcomes, specifically achievement emotions~\parencite{Pekrun2006}, self-efficacy~\parencite{Bandura1978} and interest~\parencite{Hidi2006} are affected alongside cognitive load~\parencite{Sweller1988} when using an interactive AI chatbot for physics problem solving, particularly in direct comparison to established pedagogical support systems. Specifically, the distinct effects of dynamic, interactive AI assistance versus more static aids, such as structured tiered hints or traditional textbook-style explanations, on students' affective and cognitive experiences are not yet well understood.

This study directly addresses this gap by systematically comparing the effects of AI chatbot assistance with those of tiered support and traditional textbook-style explanations. To contextualize this comparison, the following section establishes a theoretical framework by examining different forms of instructional support.

\section{Research Background} 

\subsection{Support in physics education}
A central concept in physics education is instructional scaffolding: temporary, interactive support that helps a student build on prior knowledge to complete tasks that would otherwise be out of reach~\parencite{Belland2017,Wood1976}. Crucially, effective scaffolding is not merely the provision of information but a dynamic process. Support is offered contingently - in direct response to a student's progress - and is gradually faded as the learner's competence grows~\parencite{Collins1989,Wood1976}. 

This responsive nature distinguishes scaffolding from passive instructional aids. A textbook, for example, is a static resource that cannot adapt to a learner’s immediate needs and therefore does not qualify as scaffolding~\parencite{Belland2017}. Thus, instructional support can be viewed on a spectrum of interactivity and  adaptivity. To evaluate the potential of new, highly interactive systems, it is necessary to compare them against established support methods, across this spectrum, beginning with the most traditional form of assistance.

\subsubsection{Textual Support}

Traditional textual support, most commonly embodied in textbooks, remains an important instructional tool in science classes~\parencite{OganBekiroglu2007,JonasAhrend2023}. In many science courses, such written materials are a fundamental requirement, and instructors expect students to engage with them to enhance academic performance~\parencite{French2015}. They are designed to be accessible resources that students can utilize according to their individual learning styles~\parencite{OganBekiroglu2007}. However, a significant discrepancy exists between the presumed importance of textual resources and their actual use by students. Student-reported compliance with assigned readings is varied, ranging from approximately 20\% to 70\%~\parencite{Burchfield2000}. This underutilization can be partly attributed to a perceived gap between the complexity of homework tasks and the relatively low complexity of explanations found in textbooks~\parencite{French2015}.

Learning from physics texts is an intrinsically challenging cognitive task~\parencite{Alexander1994}. The difficulty is compounded by the integration of two distinct language systems: natural language and mathematical-symbolic notation~\parencite{Alexander1994}. Furthermore, pedagogical features intended to aid comprehension, such as analogies, can sometimes increase cognitive processing demands, thereby hindering rather than helping learning~\parencite{Alexander1994}. From a constructivist perspective, a textual resource should be viewed not as a container of knowledge, but as a tool for students to construct meaning~\parencite{Cunningham1993}. Yet, the perceived authority of texts can constrain this process, as students may be less inclined to elaborate on their own thinking ~\parencite{Boxtel2000}.

Empirical work by~\Textcite{Boxtel2000} supports this, finding that student pairs with access to a textbook engaged in less collaborative elaboration compared to those without. Students often consulted the text without a specific goal, used inefficient search strategies, and failed to integrate the information with their own conceptions. Interestingly, despite the reduction in elaborative interaction, the study found no significant difference in post-test scores, suggesting that individual acquisition of factual knowledge from the text may have compensated for the lack of deeper, collaborative meaning-making. This positions textual support as a relevant, albeit potentially passive, baseline for comparison with more interactive learning support.

\subsubsection{Tiered Support}

\begin{figure}[htbp]
    \centering
    \includegraphics[width=0.5\linewidth]{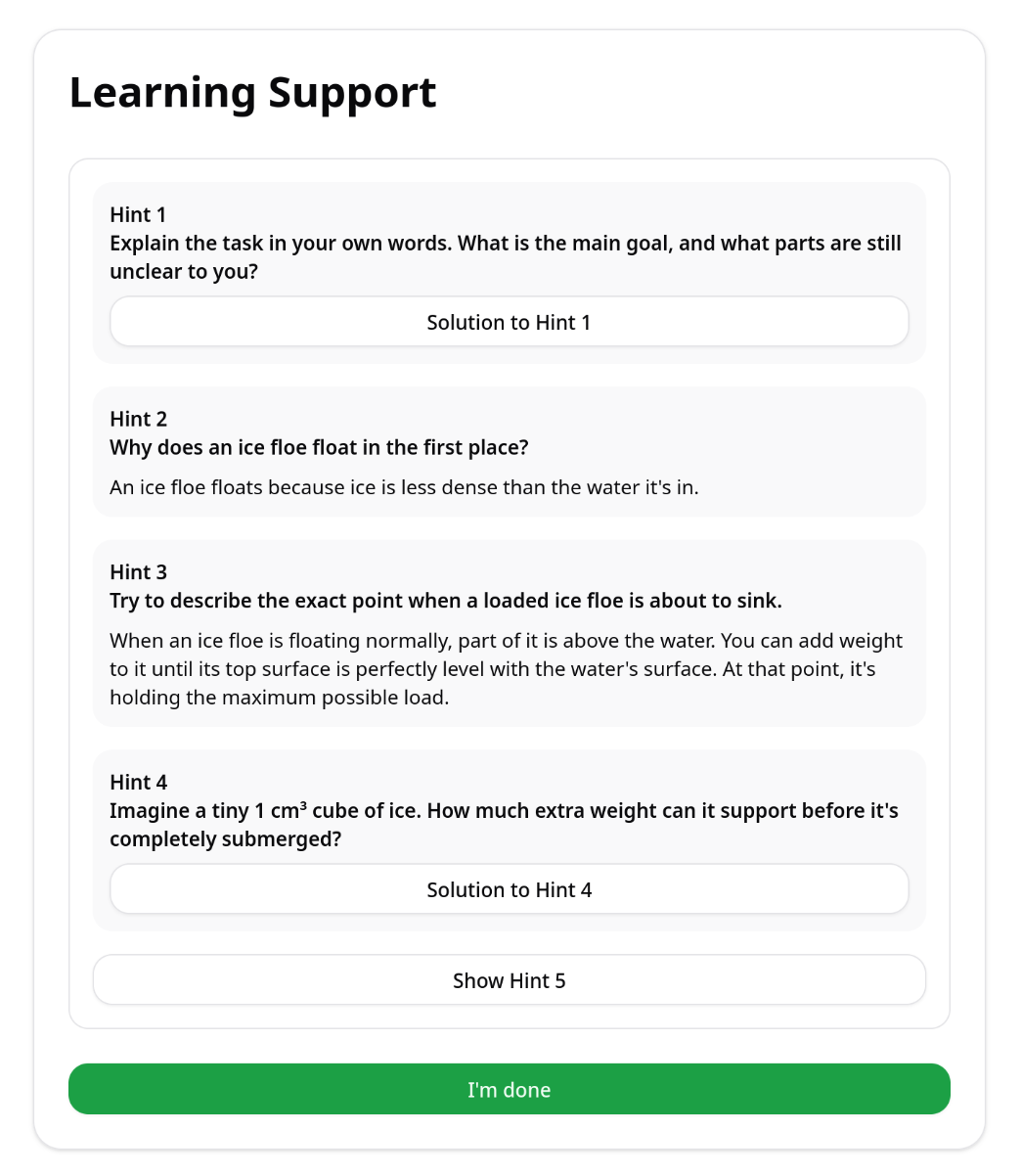}
    \caption{The digital interface for the tiered support system used in this study. Students could sequentially get a series of pre-written hints and optionally reveal their answer.}
    \label{fig:tiered-support-example}
\end{figure}

A more dynamic form of instructional assistance can be offered via scaffolding, where temporary support is provided to help a learner accomplish a task that would otherwise be out of reach~\parencite{Collins1989}. One specific implementation of this principle is ``tiered support''~\parencite[in German: ``gestufte Lernhilfen'', cf.][]{Leisen1999}. In this approach, a complex problem is accompanied by a sequence of written hints that learners can access according to their perceived needs, allowing the support to be adaptive~\parencite{StaeudelWodzinski2010}. 
These hints are designed to prompt general learning strategies, such as paraphrasing the task, focusing on relevant information, activating prior knowledge, or visualizing the problem~\parencite{FrankeBraun2008}. Typically, each hint consists of a question or prompt, with the corresponding answer made accessible in a second step~\parencite{StaeudelWodzinski2010}. The answer to the last hint is a complete sample solution that allows students to verify their answer~\parencite{FrankeBraun2008}. To illustrate this structure, Figure~\ref{fig:tiered-support-example} shows the digital interface used in this study, and the full text of the hints is provided in the supplementary materials.

The principles of tiered support are closely aligned with the extensive research on worked examples, which provide a problem statement, solution steps, and a final answer~\parencite{FrankeBraun2008}. The efficacy of both approaches can be explained by Cognitive Load Theory~\parencite[CLT, cf.][]{RenklAtkinson2003}. CLT posits that working memory has a limited capacity. When novice learners attempt to solve a problem without guidance, they often engage in inefficient search strategies (e.g., means-ends analysis) that impose a high extraneous cognitive load — effort that is not productive for learning~\parencite{Atkinson2000,RenklAtkinson2003}. By providing a clear solution path, worked examples and tiered support eliminate this counterproductive search, thereby reducing extraneous load and freeing up cognitive resources~\parencite{RenklAtkinson2003}.

Furthermore, structured methods such as tiered support help manage a material's intrinsic cognitive load. Intrinsic load is the inherent complexity of a concept, determined by the number of interacting elements that must be processed simultaneously in working memory. While this load is a fixed property of the material, instructional design can manage its impact on novices~\parencite{PollockChandlerSweller2002}. For example, segmenting a problem into smaller steps or presenting interacting elements in isolation before combining them facilitates serial processing over simultaneous processing. This strategy significantly reduces the demand on working memory~\parencite{MayerMoreno2010, PollockChandlerSweller2002}.

This process is further optimized in what is known as faded worked examples, where parts of the solution are gradually omitted, requiring the learner to fill in the gaps~\parencite{RenklAtkinson2003}. This method creates a smooth transition from fully guided example study to independent problem solving~\parencite{Atkinson2003}. Tiered support can be conceptualized as a learner-controlled fading procedure; by choosing which hints to access, students dynamically adjust the level of scaffolding, effectively creating their own incomplete example. Research has shown that such fading procedures lead to better near-transfer performance and fewer errors during learning when compared to traditional example-problem pairs~\parencite{RenklAtkinson2003}.

Empirical studies on tiered support systems have demonstrated significant motivational benefits. Students working with tiered hints reported higher feelings of competence, higher intrinsic motivation and a tendency toward greater social inclusion during partner work compared to groups receiving non-sequential help~\parencite{FrankeBraun2008,Haenze2007,SchmidtWeigand2008}. 
Furthermore, an incremental presentation of a worked example leads to higher feelings of competence and a reduced perceived cognitive load compared to presenting the solution as a whole~\parencite{SchmidtWeigand2009}. However, while incremental presentation alone improves the affective experience, the addition of strategic prompts appears necessary to significantly improve the correctness of the final solution~\parencite{SchmidtWeigand2009}. This suggests that while structured, sequential support is beneficial, the quality and function of the guidance play a key role in learning outcomes.

\subsubsection{Chatbot Support}

The third form of support, AI chatbots, offer a significant increase in interactivity and adaptivity. These conversational systems interact with users in natural language to provide answers and appropriate responses~\parencite{Huang2025,Castiglione2018}. Unlike tiered support, which follows a predefined, sequential path, AI chatbots provide dynamic and interactive scaffolding~\parencite{Huang2025}. Their primary pedagogical potential lies in their capacity for highly individualized support, which allows them to address specific student queries and guide a personalized learning process~\parencite{Kasneci2023,Neumann2024}.

The potential of AI chatbots to enhance learning has garnered considerable attention; however, a nuanced perspective is essential for interpreting the evidence. Studies indicate that while chatbots generally outperform passive control conditions, they often lag behind interventions involving human feedback or meticulously designed pedagogical approaches~\parencite{Huang2025}. Crucially, this performance gap must be understood within the context of the technologies investigated. For instance, the synthesis by \textcite{Huang2025} primarily focused on traditional, rule-based chatbots, while modern generative AI (GenAI) systems were underrepresented. This distinction is fundamental: traditional, rule-based chatbots operate on predefined scripts, whereas modern GenAI systems, powered by Large Language Models (LLMs), produce dynamic, context-aware dialogue~\parencite{Huang2025}. Importantly, these modern LLMs can be prompt-engineered -- meaning their responses and interactive behavior are meticulously shaped through specific textual system prompts -- to function as specialized pedagogical agents rather than generic conversational tools \parencite[e.g., see][]{brown2020language}.

In contrast to the mixed findings from older systems, research focusing specifically on modern LLMs demonstrates a more consistent and promising trend. For instance, a meta-analysis by \textcite{Wu2024} found that AI chatbots had a large overall effect on student learning outcomes. A recent meta-analysis on AI in STEM supports this finding, revealing large effects on student performance and learning outcomes, alongside medium effects on learning perception and higher-order thinking~\parencite{Wang2025}.

This trend is mirrored in physics education, where tutors based on Large Language Models (LLMs) show similar promise. Chatbots can serve as intelligent tutors, emulating a teacher to provide personalized guidance and feedback, or as learning tools that perform auxiliary functions like knowledge retrieval~\parencite{Wang2025}. In either capacity, these tools have been shown to improve learning performance, motivation, and student engagement.~\parencite{ChenChang2024,Kestin2024,Ng2024,Li2024,Alarbi2024,Alneyadi2023}. The benefits extend to other STEM fields, with studies in mathematics reporting significant enhancements in conceptual understanding and self-efficacy~\parencite{Canonigo2024}. Moreover, research in primary mathematics has shown that using AI to generate context-personalized tasks and materials that align with students' individual interests can signficantly increase intrinsic motivation and learning performance~\parencite{Tasdelen2025}.

However, this promising outlook is subject to critical caveats. For example, one study found that while using an LLM for a research task significantly reduced students' cognitive load, it also led to a decline in the quality of their reasoning and argumentation~\parencite{Stadler2024}. This raises crucial questions about finding the optimal balance between providing cognitive relief and ensuring students remain affectively and cognitively engaged in deep learning processes.


\section{Research Rationale} 

A central challenge in physics education is effectively addressing student heterogeneity through differentiated support~\parencite{Tomlinson2014}. This study compares three instructional tools that offer varying degrees of individualization. Textbook-style support, serving as a baseline, is a static and standardized resource that cannot adapt to individual learner needs, potentially imposing a high cognitive load~\parencite{Belland2017,Alexander1994}. An improvement is tiered support, which offers learner-directed differentiation but is too resource-intensive for educators to create for every task~\parencite{StaeudelWodzinski2010,Bosch2024}. In contrast, AI chatbots promise a significant advancement with scalable, dynamic individualization, tailoring real-time guidance to each student's specific needs~\parencite{Neumann2024}.

However, a critical gap persists in the systematic comparative evaluation of these diverse support modalities regarding their nuanced impact on learning processes. Existing research on AI in physics education has primarily focused on learning performance~\parencite{Alneyadi2023,Alarbi2024,Kestin2024}, or has been situated in gamified learning environments~\parencite{ChenChang2024} or at the university level~\parencite{Wang2024}. Other studies have explored general AI tutors~\parencite{Ng2024} or examined user experience with task-specific tutors~\parencite{Lieb2024}.

To our knowledge, no prior study has directly compared a task-specific AI chatbot explicitly designed to assist with physics problem-solving for middle school-aged students in a regular classroom context against both tiered support and traditional textual support. While previous work often emphasizes final performance outcomes, it frequently overlooks the crucial underlying cognitive processes (intrinsic and extraneous loads) and comprehensive affective experiences that shape student learning and engagement during such problem solving endeavors, a gap underscored by a recent meta-analysis calling for more research on AI chatbots' important impact on learning perception in STEM education~\parencite{Wang2025}.


A direct comparison of these support systems is necessary to move beyond isolated case studies and understand their relative strengths and weaknesses in a real-world classroom context. While theory suggests that more interactive and adaptive support should be superior, it is an open empirical question how these differences manifest in students’ learning processes. This study, therefore, provides a systematic, comparative evaluation. We examine how three distinct support types -- static textbook-style explanations, tiered support, and a dynamic AI chatbot -- impact two critical dimensions of the learning experience: the cognitive burden students face during the task and their concurrent affective and motivational states. By doing so, we seek to generate practical, evidence-based insights to help educators choose and design support tools that enhance both learning efficiency and student well-being. Thus, this investigation is guided by the following research questions:

\begin{enumerate}[leftmargin=15mm]
     \item[\bf RQ1:] To what extent do AI chatbot support, tiered support, and traditional textual support differ in their effects on students' perceived cognitive load during physics problem solving?
     
    \item[\bf RQ2:] To what extent do AI chatbot support, tiered support, and traditional textual support differ in their effects on students' interest, self-efficacy, and achievement emotions (hope, hopelessness, enjoyment) during physics problem solving?

\end{enumerate}

\section{Materials \& Methods} 

\subsection{Study design and sample}

We conducted a quasi-experimental\footnote{Cluster randomization was performed at class level.} field study using a pre-post-test design to compare the effects of different learning support systems on student motivation and cognitive load during physics problem solving. The final sample consisted of 273 ninth-grade students who met two inclusion criteria: they had completed all questionnaires and explicitly confirmed their use of the support system for problem-solving during the study. The study was conducted within the students' classrooms and followed a standardized protocol to ensure consistency across groups. Students were instructed to work silently and individually.

During the intervention phase, each student individually solved a physics task on buoyancy, adapted from~\textcite{FrankeBraun2008}. The task was presented on a paper worksheet, and students were allotted 20 minutes for completion and an additional 5 minutes to compare their work with a sample solution. Each worksheet also contained a brief guide for the accompanying support tool, which offered practical instructions for its use and guidance on applying metacognitive strategies. Students recorded their answers and reasoning directly on the worksheet. 

All participants performed the same core task, but the learning support they received varied across three experimental conditions illustrated in Table~\ref{tab:conditions}.

\begin{table}[htbp]
\caption{Overview of Experimental Conditions.}
\label{tab:conditions}
\centering
\begin{tabular}{l l c}
\hline\hline
\textbf{Group} & \textbf{Description of Support Condition} & \textbf{N} \\
\hline
\text{G1: Chatbot} &
\begin{minipage}[t]{0.6\textwidth}
Participants interacted with a custom-configured AI chatbot for task-related guidance and feedback. The chatbot was designed to promote independent problem solving and conceptual understanding.
\end{minipage} &
103 \\[4ex]
\text{G2: Tiered Support} &
\begin{minipage}[t]{0.6\textwidth}
Participants were provided with a series of structured, step-by-step hints on a website, which could be revealed sequentially~\parencite[adapted from][]{FrankeBraun2008}.
\end{minipage} &
64 \\[4ex]
G3: \begin{tabular}[t]{@{}l@{}}Textual Support \\ (Control)\end{tabular} &
\begin{minipage}[t]{0.6\textwidth}
Participants received a static digital textbook-style page, which contained all the information required to solve the problem.
\end{minipage} &
101 \\
\hline
\end{tabular}
\end{table}

\begin{figure}[htbp]
  \centering
\includegraphics[width=\textwidth]{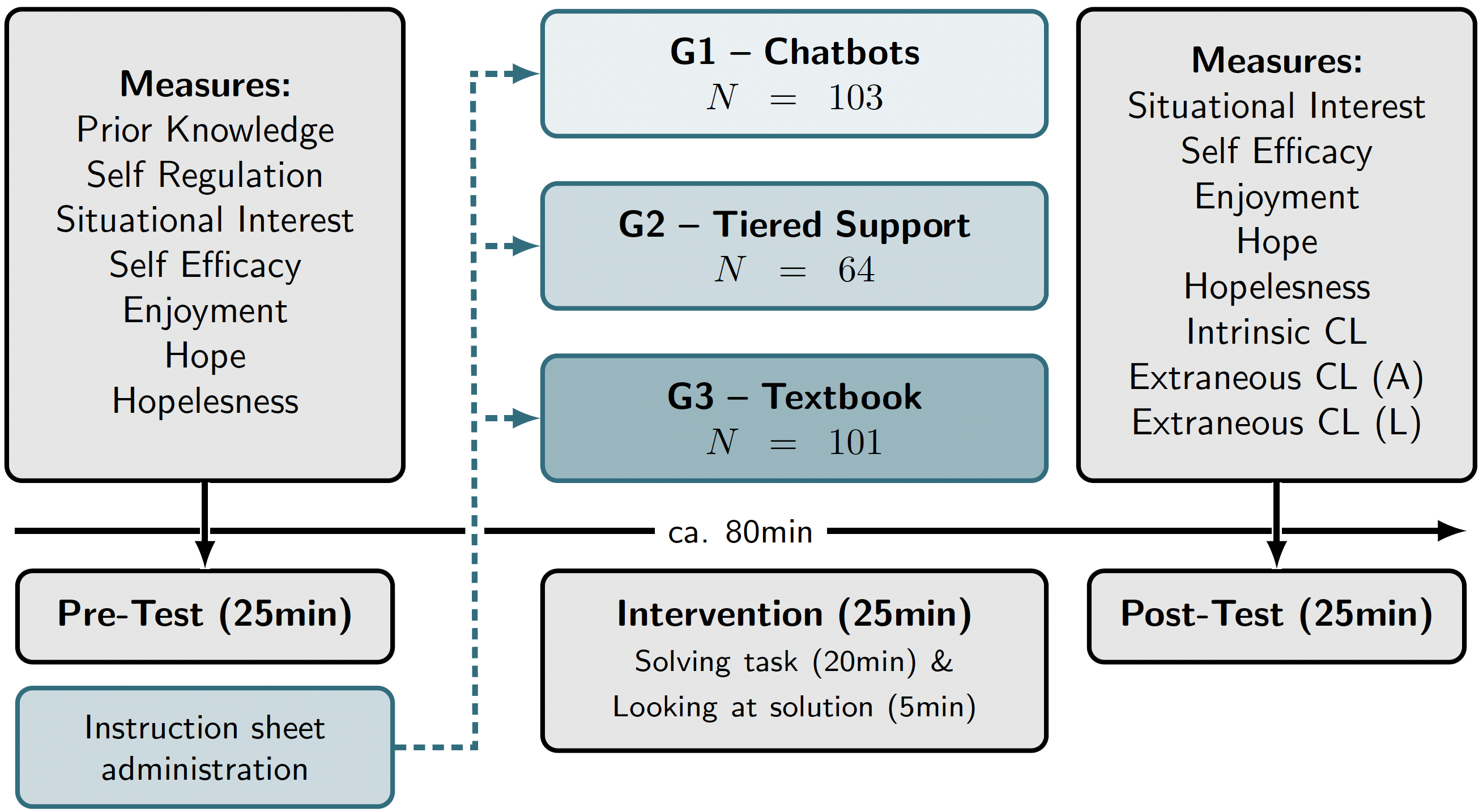}
  \caption{Overview of the study design, including the three different groups and the corresponding sample sizes as well as the assessed variables and the dedicated time slots.}
  \label{fig:Study-Design}
\end{figure}

To assess the effects of the intervention, data was collected via questionnaires administered directly before (pre-test) and after (post-test) the intervention. Individual student responses were linked across the two time points using an anonymous identifier. The dependent variables were measures of interest, self-efficacy, cognitive load, achievement emotions (hope, hopelessness, enjoyment). To account for potential confounding factors, the following control variables were assessed: students' self-regulation skills and prior knowledge in physics. A detailed overview of all measured variables, including their abbreviations, measurement times, and scale properties, can be found in Table~\ref{tab:instrument}. The study design is visualized in Figure \ref{fig:Study-Design}.

\subsection{Variable Selection}

To investigate our research questions, we developed a questionnaire to assess key affective and cognitive variables before and after the intervention. The selection of these variables was guided by established theoretical frameworks from educational psychology and physics education research. 

To isolate the effects of the different support conditions, we controlled for two pre-existing student characteristics: First, we measured prior conceptual understanding of buoyancy, (a) to check for differences in students' prior knowledge across the comparison groups, and (b) since this variable can be regarded a key determinant of learning and is known to correlate with motivation, achievement emotions and cognitive load~\parencite{Pintrich1990,Pekrun2024,Sweller2019,Dong2020}. Because the effective use of learner-controlled tools like AI chatbots and tiered support likely depends on a student's ability to monitor their understanding and seek help appropriately, we included students' trait self-regulation (SR) as a further control variable to account for its potential influence on the outcomes~\parencite{Kelly2015}. The rationale for selecting our primary dependent variables is detailed in the remainder of this section.

Our first research question addressed the cognitive demands of the task, which we assessed through the lens of Cognitive Load Theory (CLT)~\parencite{Sweller1988,vanMerrienboer2005}. We measured both intrinsic and extraneous cognitive load. We hypothesized that the chatbot and tiered support interventions would result in lower extraneous load compared to the static textual support. We further posited that this reduction is achieved through distinct mechanisms:
In the Tiered Support condition, by presenting solution steps sequentially, the intervention leverages the isolated elements effect, reducing the load of managing many interacting elements at once~\parencite{PollockChandlerSweller2002}.
In the chatbot condition, the human-AI dyad leverages the collective working memory effect, where the AI partner offloads the cognitive burden of structuring the problem solving process~\parencite{Kirschner2009}. Furthermore, the chatbot’s Socratic dialogue is designed to elicit self-explanation, a process that converts potentially extraneous effort into productive cognitive processing~\parencite{Renkl1998,Sweller2019}.

Our second research question examined the impact of the support conditions on motivational outcomes. Accordingly, we measured situational interest, as it is a powerful influence on attention and the depth of learning~\parencite{Hidi2006}. We hypothesized that the interactive nature of the AI chatbot, a known trigger for situational interest, would be particularly effective in engaging students. In addition, we assessed self-efficacy, a student's belief in their capability to succeed, which is a critical determinant of effort and perseverance in difficult physics problems~\parencite{Iwuanyanwu2022}. Grounded in theory by \textcite{Bandura1978}, we posited that interactive support could enhance self-efficacy by providing mastery experiences and verbal persuasion. This aligns with findings that scaffolding systematic problem solving builds self-efficacy in physics~\parencite{Maries2023}. Finally, to capture a more nuanced picture of the affective experience, we measured key achievement emotions. Based on recent findings, we focused on hope, hopelessness, and enjoyment, as these emotions are highly predictive of academic outcomes and are linked to the cognitive resources students deploy during a task~\parencite{Pekrun2006}.

\subsection{Instrument Development}

The questionnaire used in this study was developed using established and research-validated scales to measure the variables of interest. Scales originally developed in English, were translated into German, with meticulous attention to preserving their conceptual meaning. To ensure relevance to the specific physics context, namely buoyancy, the wording of some items and the introductory text for each scale were systematically adapted. These adapted items were then reviewed by two physics education experts to ensure their meaning was preserved. For consistency across measures and to simplify the response process for students, all variables were assessed using a uniform 5-point rating scale, ranging from 1 (``Does not apply at all'') to 5 (``Fully applies'').  A comprehensive overview of all constructs, their corresponding scales, source references, and estimates of internal consistency (Cronbach's $\alpha$) is provided in Table~\ref{tab:instrument}. 

\subsection{Materials}

The final materials for all three conditions -- the textual support page, the tiered support task, and the chatbot's system prompt -- are provided in the supplementary materials alongside the buoyancy task. Here, we provide an overview of the development of the support materials. 

\subsubsection{Textual Support Development}

The textual support material for the control group was developed through a systematic process of design including expert feedback. Its creation was guided by a set of explicit design principles derived from textbook research, ensuring the material featured: (1) a clear structure and didactically appropriate content for the 9th-grade target group; (2) precise, linguistically correct text; and (3) high-quality, supportive visualizations~\parencite{Devetak2013, Fuchs2014}. To verify its pedagogical quality and content validity, the resulting digital A4-page was reviewed and approved by two experts in physics education research. The final, self-contained document provided all necessary information for a student to solve the problem.

\subsubsection{Tiered Support Development}

For the scaffolding group, the intervention utilized a specific tiered support (gestufte Lernhilfen) task taken directly from the materials developed by~\Textcite{FrankeBraun2008}. While the original design was intended for collaborative pair work, the materials were modified for individual problem solving to align with our study's protocol. The support was implemented digitally on a custom website where students could sequentially reveal hints on a step-by-step basis at their own discretion. The structure mirrored the original design: each tier consisted of a prompt or question that students could reveal, followed by a corresponding answer that could be unlocked in a second step. This learner-controlled process concluded with a final tier containing the complete model solution, allowing students to verify their work.

\subsubsection{Chatbot Development}

The AI chatbot used in this study was developed using OpenAI's o3-mini model accessed via the Fobizz platform\footnote{Accessible via \url{https://fobizz.com/en/}}. Its pedagogical behavior was meticulously prompt-engineered through a rigorous iterative refinement cycle, a methodology established in the development of educational technology~\parencite{Bassner2024,Kestin2024,Wan2024}. During each cycle, the chatbot's responses to simulated student inputs were evaluated against three key pedagogical criteria derived from research on expert tutoring: (1) its ability to accurately assess a student's current understanding; (2) its capacity to pose guiding, reflective questions rather than providing direct corrections; and (3) its adherence to a ``build on what's right'' principle by first affirming correct elements of a student's reasoning before prompting them to reconsider flawed parts~\parencite{Lepper1997,Lemmrich2024}. This iterative process was continued until the chatbot consistently met these standards, resulting in the final system prompt (included in the supplementary materials), which instructed the chatbot to act as a supportive tutor, primarily designed to guide students toward independent problem solving and conceptual understanding.

\subsection{Data Analysis}

We first examined the relationship between all variables (cf. Table \ref{tab:instrument}), separated by pre- and post-test, using Spearman’s rank-order correlation coefficient $\rho$ -- which is suitable for correlating ordinal variables -- in conjunction with the widely accepted classification into weak ($|\rho|<0.20$), medium ($0.20\leq |\rho|<0.30$) and strong ($|\rho|\geq 0.30$) correlations according to~\textcite{Hemphill03}. Since the correlation analysis revealed strong intercorrelations among the variables (cf. Table \ref{tab:correlations-pre}), a one-way MANOVA was conducted using Wilk's $\Lambda$ (which is suitable in the case of roughly similar sample sizes, normality of the data and homogeneity of variance, cf. \cite{Mood19}) to check for statistical significance of group differences in pre-test scores. We found no statistically significant multivariate differences [$F(16,530)=0.90$, Wilk's $\Lambda =0.948$, $p=0.565$]. To further substantiate this result, we performed separate one-way ANOVAs for each pre-test variable, which likewise revealed no significant differences. Given that the pre-test differences were statistically negligible, we proceeded without including pre-test scores as covariates in subsequent analyses. For the post-test data, we again employed a one-way MANOVA due to the strong correlations among the dependent variables (cf. Table \ref{tab:correlations-post}). Following the detection of a significant multivariate effect [$F(16,528)=3.83$, Wilk's $\Lambda=0.80$, $p<0.001$], we conducted separate one-way ANOVAs to identify which post-test variables contributed to the overall group differences. As a measure of effect size, we used partial eta squared ($\eta_p^2$), alongside the commonly used thresholds for small ($\eta_p^2<0.06$), medium ($0.06\leq\eta_p^2<0.14$) and large ($0.14\leq\eta_p^2$) effects~\parencite{Cohen88}. Lastly, we used Tukey-Kramer post-hoc tests to check for significant differences between the groups, which include the Tukey Honestly Significant Difference (HSD) correction to prevent accumulating type I errors when conducting all pairwise group comparisons \parencite{Abdi10}. To gauge the effect of these pairwise comparisons, we used Cohen's d and the established ranges of small ($d<0.5$), medium ($0.5\leq d<0.8$) and large ($0.8\leq d$) effect sizes~\parencite{Cohen88}. Prerequisites of analyses of variance were checked by following~\textcite{Lin17} and using Levene's Test to ensure homogeneity of variance~\parencite{Levene60}, Box's M Test to ensure homogeneity of covariances~\parencite{Box53} as well as Shaprio-Wilk Test and Mardia's Test to ensure univariate and multivariate normality of the data~\parencite{Shapiro65,Mardia75}. Linearity and multicollinearity were tested visually using scatter plots as well as numerically using correlation tables (cf. Table~\ref{tab:correlations-pre} and Table~\ref{tab:correlations-post}).


To complement the MANOVA results and provide a more holistic picture of the data, we further report general descriptive statistics, including response distribution, mean values, standard deviation (SD) as well as median values for all indicators, split by intervention group. As mentioned above, we additionally provide Cronbach's $\alpha$ as a measure of internal consistency for each scale, where values above 0.70 are considered acceptable~\parencite{Taber18}.

The data analysis was conducted using R (version 4.4.2) and its packages pwr (version 1.3-0), mvn (version 5.9), biotools (version 4.3), effectsize (version 1.0.1), emmeans (version 1.11.1), and heplots (1.7.5).

\section{Results}
Table~\ref{tab:instrument} provides an overview in terms of scale length and Cronbach's $\alpha$. While most scales demonstrated good internal consistency ($\alpha$ > 0.70), the brief scales for intrinsic cognitive load ($\alpha$ = 0.58) and extraneous cognitive load of the exercise ($\alpha$ = 0.61) exhibited lower internal consistency, alongside the prior knowledge test ($\alpha=0.63$). 

\begin{table}[htbp]
  \caption{Overview of all assessed variables in terms of construct abbreviation, scale length, measurement times and internal consistency expressed via Cronbach's $\alpha$. If scales have been administered at both pre- and post-test points in time, two values for Cronbach's $\alpha$ are reported.}
  \label{tab:instrument}
  \begin{tabular}{lcccl}
  \toprule
  Variable  (Abbr., \# Items) &  Pretest & Posttest & Cronbach's $\alpha$& Reference\\
  \hline
  Prior Knowledge (PK, 9) & $\checkmark$ & $\times$ & 0.63& \cite{Zenger2022} \\
  Self Regulation (SR, 9) & $\checkmark$ & $\times$ & 0.73 & \cite{Pintrich1990} \\
  Situational Interest (SI, 6) & $\checkmark$ & $\checkmark$ & 0.81 \& 0.85 & \cite{Woithe2020} \\
  Self Efficacy (SE, 7) & $\checkmark$ & $\checkmark$ & 0.88 \& 0.90 & \cite{Duncan2015} \\
  Enjoyment (EE, 4) & $\checkmark$ & $\checkmark$ & 0.79 \& 0.88 & \cite{Bieleke2021} \\ 
  Hope (EH, 4) & $\checkmark$ & $\checkmark$ & 0.88 \& 0.89 & \cite{Bieleke2021} \\ 
  Hopelessness (EHL, 4) & $\checkmark$ & $\checkmark$ & 0.84 \& 0.82 & \cite{Bieleke2021} \\ 
   
  Intrinsic CL (ICL, 2) & $\times$ & $\checkmark$ & 0.58 & \cite{Klepsch2017} \\ 
  Extraneous CL Exercise (ECLA, 3) & $\times$ & $\checkmark$ & 0.61 & \cite{Klepsch2017} \\ 
  Extraneous CL Support (ECLL, 3) & $\times$ & $\checkmark$ & 0.81 & \cite{Klepsch2017} \\ 
 
  \bottomrule
  \end{tabular}
\end{table}

The relationship between all variables in terms of Spearman's $\rho$ is provided in Table~\ref{tab:correlations-pre} for the pre-test and in Table~\ref{tab:correlations-post} for the post-test. As mentioned in the Data Analysis section, all variables in both tests are significantly correlated ($p<0.001$). In the pre-test, their absolute values range from the weak correlation $\rho=0.17$ (between prior knowledge and self regulation) to the strong correlation $\rho=0.70$ (between hope and self-efficacy) while in the post-test they exhibit a similar range from $\rho=0.17$ (between self-efficacy and intrinsic CL) to $\rho=0.71$ (between hope and enjoyment).

\begin{table}[htbp]
  \caption{Spearman's rank-order correlation coefficient $\rho$ for all pairs of pre-test variables.}
  \label{tab:correlations-pre}
  \begin{tabular}{c|ccccccc}
  \toprule
  & DI & SR & SE & EE & EH & EHL & PK \\
  \midrule
  DI & -- &&&&&&\\
  SR & $0.50^{\star\star\star}$ &--&&&&& \\
  SE & $0.61^{\star\star\star}$ & $0.53^{\star\star\star}$ &--&&&& \\
  EE & $0.62^{\star\star\star}$ & $0.50^{\star\star\star}$ & $0.48^{\star\star\star}$ &--&&& \\
  EH & $0.57^{\star\star\star}$ & $0.54^{\star\star\star}$ & $0.70^{\star\star\star}$ & $0.63^{\star\star\star}$ &--&& \\
  EHL & $-0.40^{\star\star\star}$ & $-0.44^{\star\star\star}$ & $-0.52^{\star\star\star}$ & $-0.28^{\star\star\star}$ & $-0.49^{\star\star\star}$ &--&\\
  PK & $0.21^{\star\star\star}$ & $0.17^{\star\star\star}$ & $0.27^{\star\star\star}$ & $0.27^{\star\star\star}$ & $0.27^{\star\star\star}$ & $-0.24^{\star\star\star}$ &--\\
  \hline
  \bottomrule
  \end{tabular}
  \tablenote{Signif. codes:  $p<0.001$ ‘$^{\star\star\star}$’, $p<0.01$ ‘$^{\star\star}$’, $p<0.05$ ‘$^{\star}$’}
\end{table}

\begin{table}[htbp]
  \caption{Spearman's rank-order correlation coefficient $\rho$ for all pairs of post-test variables.}
  \label{tab:correlations-post}
  \begin{tabular}{c|cccccccc}
  \toprule
  & EE & EH & EHL & ICL & ECLA & ECLL & SI & SE \\
  \midrule
  EE & -- &&&&&&&\\
  EH & $0.71^{\star\star\star}$ &--&&&&&& \\
  EHL & $-0.42^{\star\star\star}$ & $-0.47^{\star\star\star}$ &--&&&&& \\
  ICL & $-0.22^{\star\star\star}$ & $-0.25^{\star\star\star}$ & $0.37^{\star\star\star}$ &--&&&& \\
  ECLA & $-0.27^{\star\star\star}$ & $-0.26^{\star\star\star}$ & $0.45^{\star\star\star}$ & $0.45^{\star\star\star}$ &--&&& \\
  ECLL & $-0.38^{\star\star\star}$ & $-0.30^{\star\star\star}$ & $0.44^{\star\star\star}$ & $0.37^{\star\star\star}$ & $0.51^{\star\star\star}$ &--&&\\
  SI & $0.69^{\star\star\star}$ & $0.60^{\star\star\star}$ & $-0.29^{\star\star\star}$ & $-0.06$ & $-0.22^{\star\star\star}$ & $-0.31^{\star\star\star}$ &--&\\
  SE & $0.50^{\star\star\star}$ & $0.62^{\star\star\star}$ & $-0.49^{\star\star\star}$ & $-0.17^{\star\star}$ & $-0.19^{\star\star}$ & $-0.23^{\star\star\star}$ & $0.51^{\star\star\star}$ &--\\
  \hline
  \bottomrule
  \end{tabular}
  \tablenote{Signif. codes:  $p<0.001$ ‘$^{\star\star\star}$’, $p<0.01$ ‘$^{\star\star}$’, $p<0.05$ ‘$^{\star}$’}
\end{table}

Since the three groups did not differ in any pre-test variable, only the post-test variables were considered for further analysis (cf. Section Data Analysis). Descriptive statistics for all post-test variables are thus provided in Table~\ref{tab:descriptives-1} and Table~\ref{tab:descriptives-2}, including means, standard deviations, medians, minima, and maxima.

\begin{table}[htbp]
  \caption{Descriptive statistics for variables EE (a), EH (b), EHL (c), and ICL (d), split by group -- Chatbots (G1), Tiered Support (G2), Textbook (G3).}
  \centering

  \begin{subtable}[b]{0.45\textwidth}
    \centering
       \caption{Descriptives for the variable \textit{Enjoyment} (EE) for each group.}
  \label{tab:descriptives-1-ee} 
  \begin{tabular}{cccccc}
  \toprule
  Group & Mean & SD & Median & Min & Max \\
  \midrule
  G1 & 2.65 & 0.93 & 2.50 & 1 & 5\\
  G2 & 2.47 & 0.88 & 2.50 & 1 & 5\\
  G3 & 2.20 & 0.89 & 2.25 & 1 & 5\\
  \hline
  \bottomrule
  \end{tabular}

  \end{subtable}
  \hfill
  \begin{subtable}[b]{0.45\textwidth}
    \centering
        \caption{Descriptives for the variable \textit{Hope} (EH) for each group.}
    \label{tab:descriptives-1-eh}
  \begin{tabular}{cccccc}
  \toprule
  Group & Mean & SD & Median & Min & Max \\
  \midrule
  G1 & 2.62 & 0.87 & 2.75 & 1 & 5\\
  G2 & 2.45 & 0.87 & 2.50 & 1 & 5\\
  G3 & 2.14 & 0.84 & 2.25 & 1 & 5\\
  \hline
  \bottomrule
  \end{tabular}

  \end{subtable}

  \vspace{0.5cm}

  \begin{subtable}[b]{0.45\textwidth}
    \centering
    \caption{Descriptives for the variable \textit{Hopelessness} (EHL) for each group.}
    \label{tab:descriptives-1-ehl}
   \begin{tabular}{cccccc}
  \toprule
  Group & Mean & SD & Median & Min & Max \\
  \midrule
  G1 & 2.60 & 0.97 & 2.50 & 1 & 5\\
  G2 & 2.94 & 1.00 & 3.00 & 1 & 5\\
  G3 & 3.15 & 0.94 & 3.13 & 1 & 5\\
  \hline
  \bottomrule
  \end{tabular}
  \end{subtable}
  \hfill
  \begin{subtable}[b]{0.45\textwidth}
    \centering
       \caption{Descriptives for the variable \textit{Intrinsic CL} (ICL) for each group.}
    \label{tab:descriptives-1-icl} 
  \begin{tabular}{cccccc}
  \toprule
  Group & Mean & SD & Median & Min & Max \\
  \midrule
  G1 & 3.18 & 0.82 & 3.00 & 1 & 5\\
  G2 & 3.27 & 0.75 & 3.25 & 1 & 5\\
  G3 & 3.74 & 0.77 & 4.00 & 1 & 5\\
  \hline
  \bottomrule
  \end{tabular}

  \end{subtable}

  \label{tab:descriptives-1}
\end{table}

\begin{table}[htbp]

  \caption{Descriptive statistics for variables ECLA (a), ECLL (b), SI (c), and SE (d), split by group -- Chatbots (G1), Tiered Support (G2), Textbook (G3).}
  \centering

  \begin{subtable}[b]{0.45\textwidth}
    \centering
       \caption{Descriptives for the variable \textit{Extraneous CL Exercise} (ECLA) for each group.}
  \label{tab:descriptives-2-ecla} 
  \begin{tabular}{cccccc}
  \toprule
  Group & Mean & SD & Median & Min & Max \\
  \midrule
  G1 & 2.92 & 0.76 & 3.00 & 1 & 5\\
  G2 & 2.84 & 0.70 & 3.00 & 1 & 5\\
  G3 & 3.42 & 0.75 & 3.33 & 1 & 5\\
  \hline
  \bottomrule
  \end{tabular}

  \end{subtable}
  \hfill
  \begin{subtable}[b]{0.45\textwidth}
    \centering
        \caption{Descriptives for the variable \textit{Extraneous CL Support} (ECLL) for each group.}
    \label{tab:descriptives-2-ecll}
  \begin{tabular}{cccccc}
  \toprule
  Group & Mean & SD & Median & Min & Max \\
  \midrule
  G1 & 2.69 & 0.93 & 2.67 & 1 & 5\\
  G2 & 2.74 & 0.98 & 2.67 & 1 & 5\\
  G3 & 3.26 & 0.92 & 3.00 & 1 & 5\\
  \hline
  \bottomrule
  \end{tabular}

  \end{subtable}

  \vspace{0.5cm}

  \begin{subtable}[b]{0.45\textwidth}
    \centering
    \caption{Descriptives for the variable \textit{Situational Interest} (SI) for each group.}
    \label{tab:descriptives-2-si}
  \begin{tabular}{cccccc}
  \toprule
  Group & Mean & SD & Median & Min & Max \\
  \midrule
  G1 & 2.46 & 0.83 & 2.50 & 1 & 5\\
  G2 & 2.21 & 0.77 & 2.33 & 1 & 5\\
  G3 & 2.11 & 0.77 & 2.17 & 1 & 5\\
  \hline
  \bottomrule
  \end{tabular}
  \end{subtable}
  \hfill
  \begin{subtable}[b]{0.45\textwidth}
    \centering
       \caption{Descriptives for the variable \textit{Self Efficacy} (SE) for each group.}
    \label{tab:descriptives-2-se} 
  \begin{tabular}{cccccc}
  \toprule
  Group & Mean & SD & Median & Min & Max \\
  \midrule
  G1 & 2.89 & 0.77 & 3.00 & 1 & 5\\
  G2 & 2.72 & 0.72 & 2.86 & 1 & 5\\
  G3 & 2.55 & 0.81 & 2.64 & 1 & 5\\
  \hline
  \bottomrule
  \end{tabular}

  \end{subtable}

  \label{tab:descriptives-2}
\end{table}

\begin{figure}[htbp]
  \centering
  
  \begin{subfigure}[b]{0.45\textwidth}
    \centering
    \includegraphics[width=\textwidth]{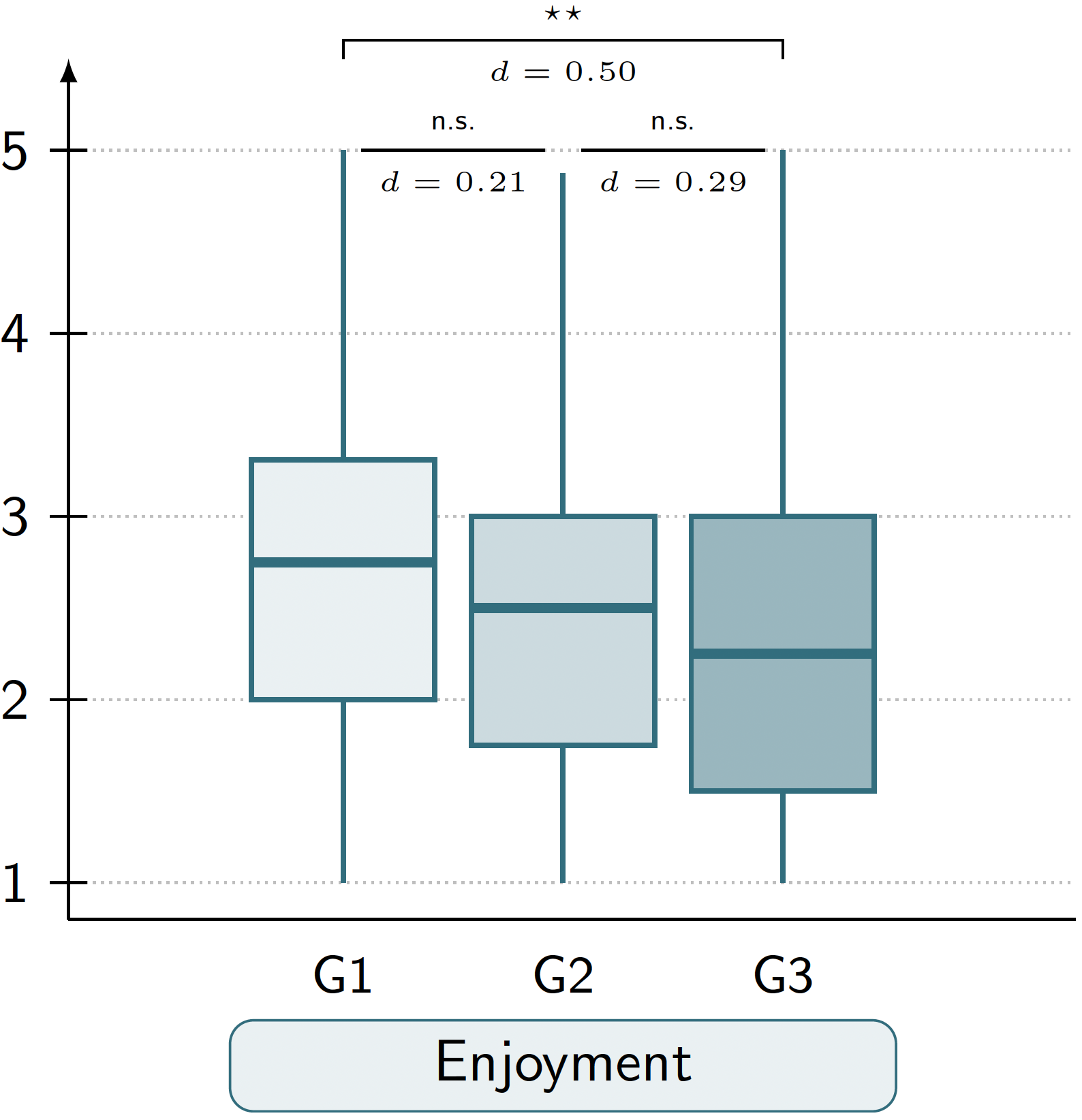}
    \caption{Boxplots for the variable \textit{Enjoyment} (EE) for each group.}
    \label{fig:boxplots-1-ee}
  \end{subfigure} \begin{subfigure}[b]{0.45\textwidth}
    \centering
    \includegraphics[width=\textwidth]{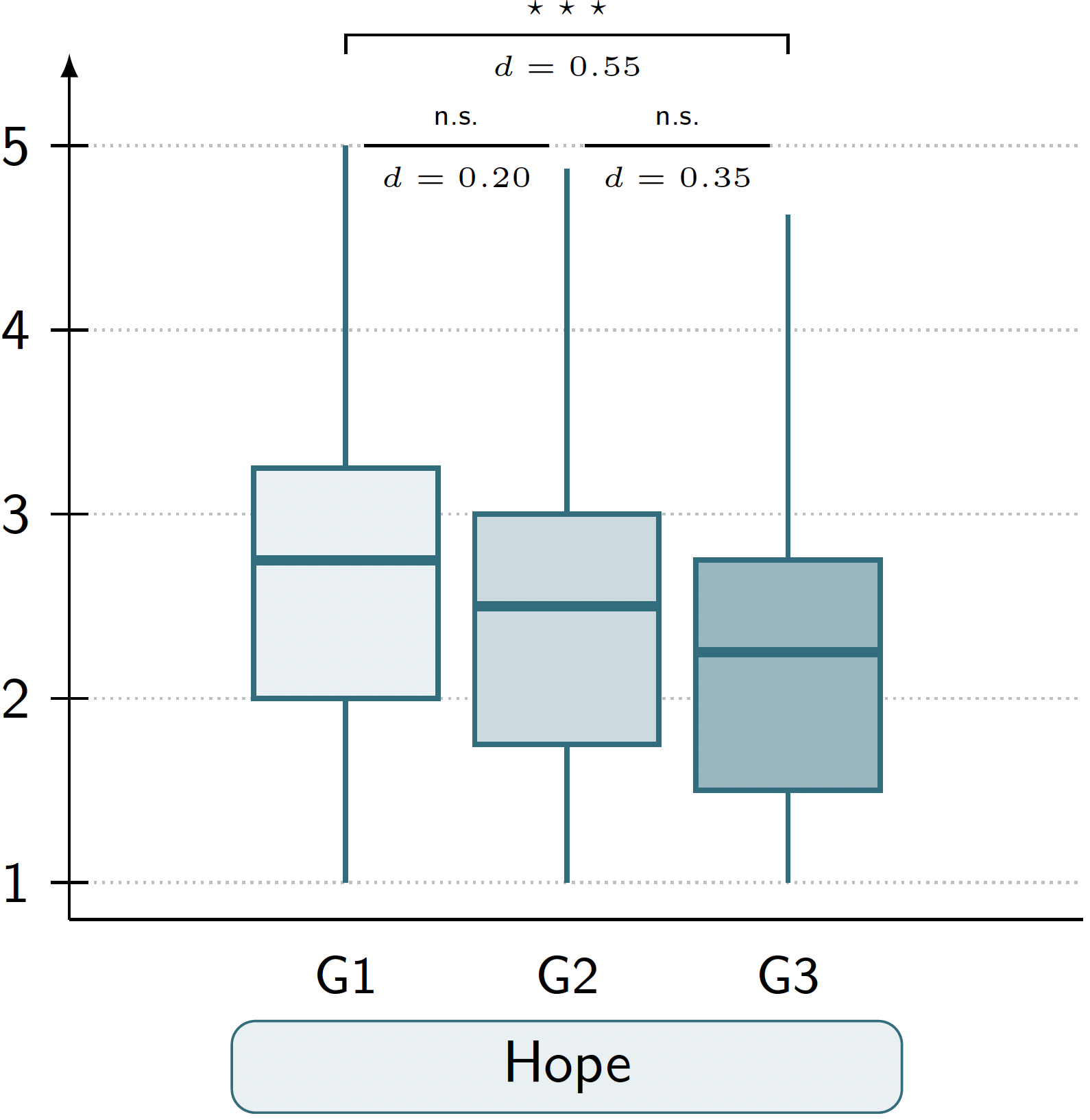}
    \caption{Boxplots for the variable \textit{Hope} (EH) for each group.}
    \label{fig:boxplots-1-eh}
  \end{subfigure}

  \vspace{0.5cm} 

  \begin{subfigure}[b]{0.45\textwidth}
    \centering
    \includegraphics[width=\textwidth]{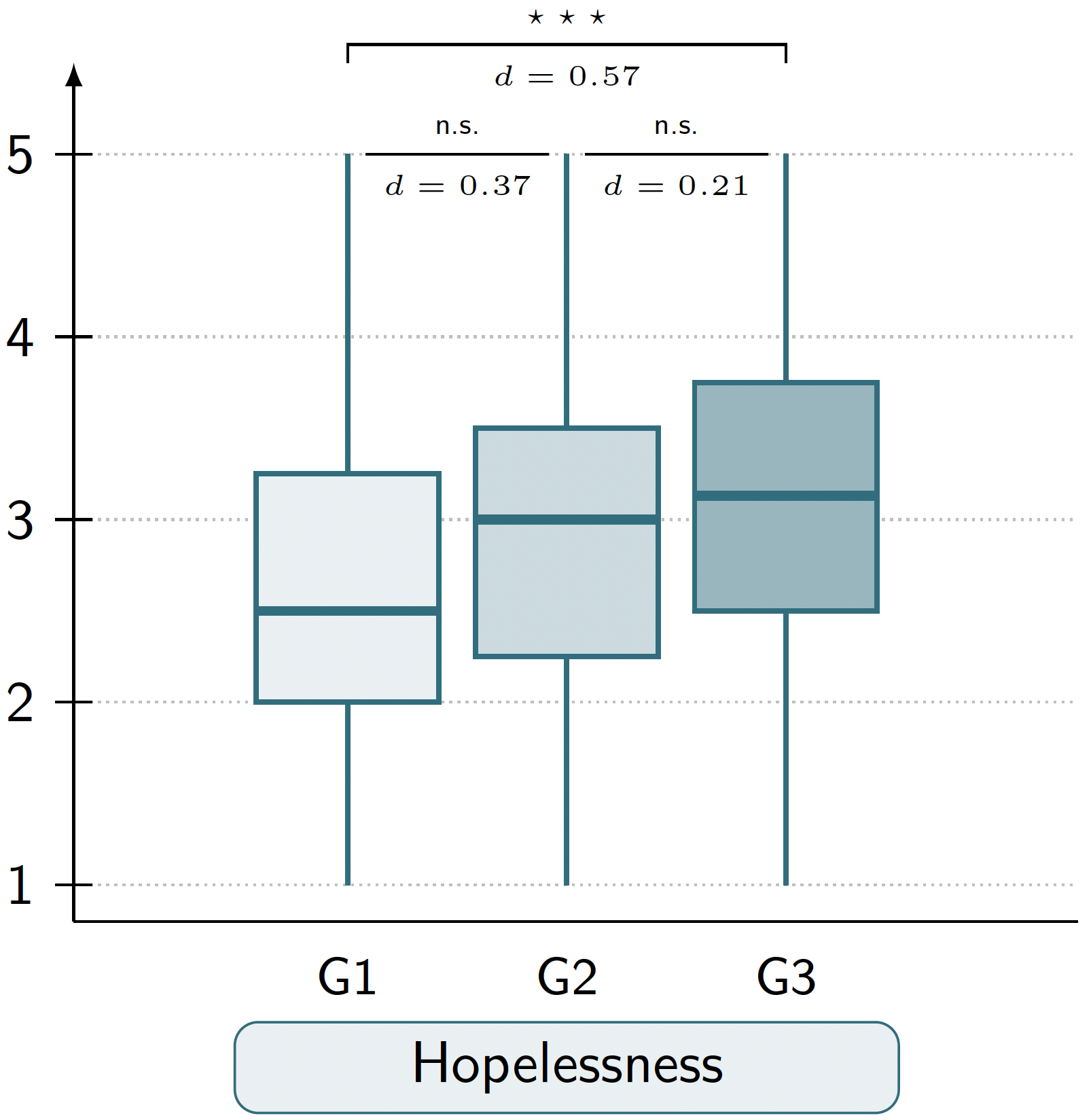}
    \caption{Boxplots for the variable \textit{Hopelessness} (EHL) for each group.}
    \label{fig:boxplots-1-ehl}
  \end{subfigure} \begin{subfigure}[b]{0.45\textwidth}
    \centering
    \includegraphics[width=\textwidth]{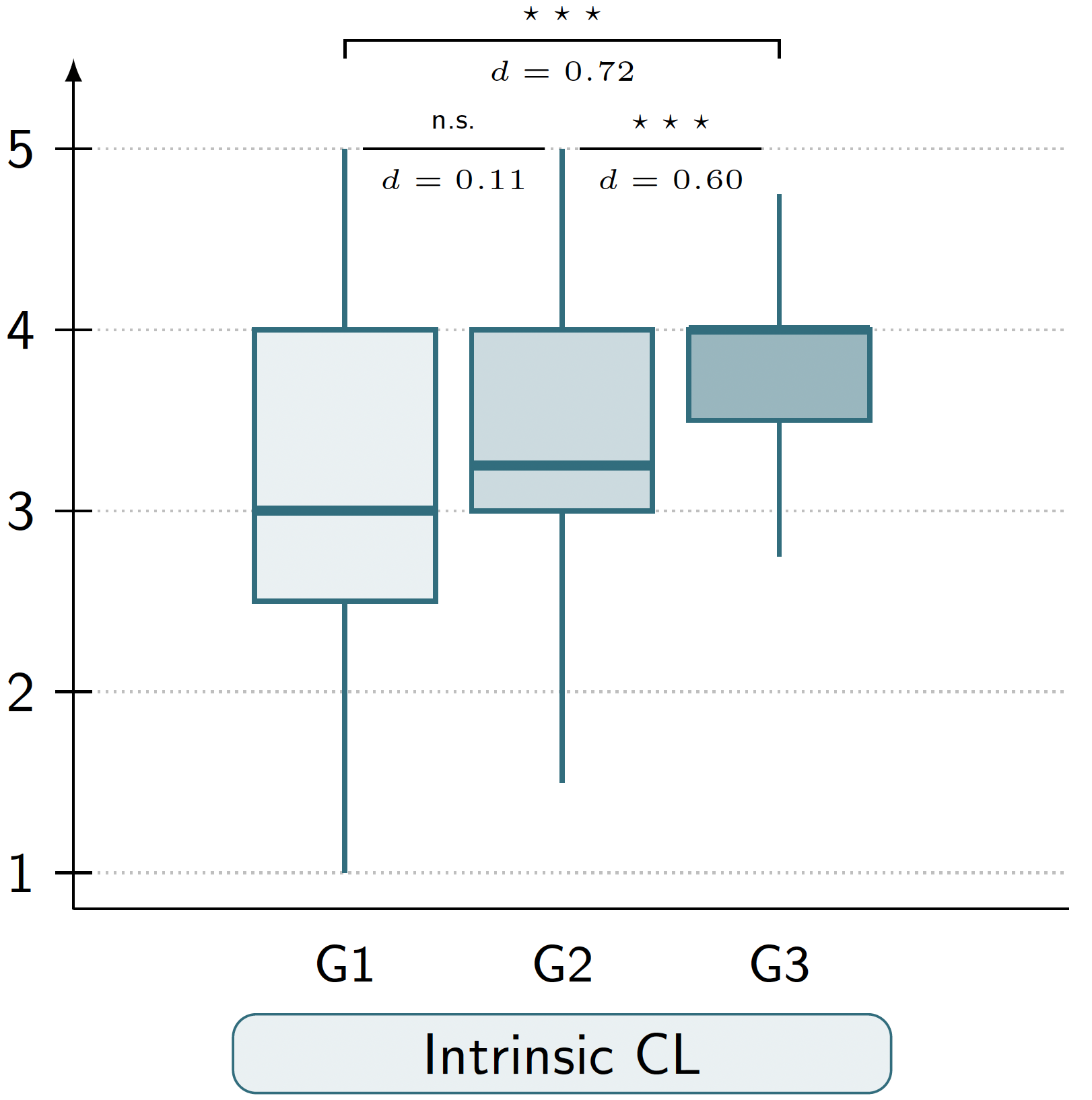}
    \caption{Boxplots for the variable \textit{Intrinsic CL} (ICL) for each group.}
    \label{fig:boxplots-1-icl}
  \end{subfigure}

  \caption{Boxplots for the variables EE (subfigure a), EH (subfigure b), EHL (subfigure c), and ICL (subfigure d), split by group -- Chatbots (G1), Tiered Support (G2), Textbook (G3). In addition, Cohen's d as a measure of effect size is reported alongside the respective significance level of the ANOVA F-test for each comparison.}
  \label{fig:boxplots-1}
\end{figure}

\begin{figure}[htbp]
  \centering
  
  \begin{subfigure}[b]{0.45\textwidth}
    \centering
    \includegraphics[width=\textwidth]{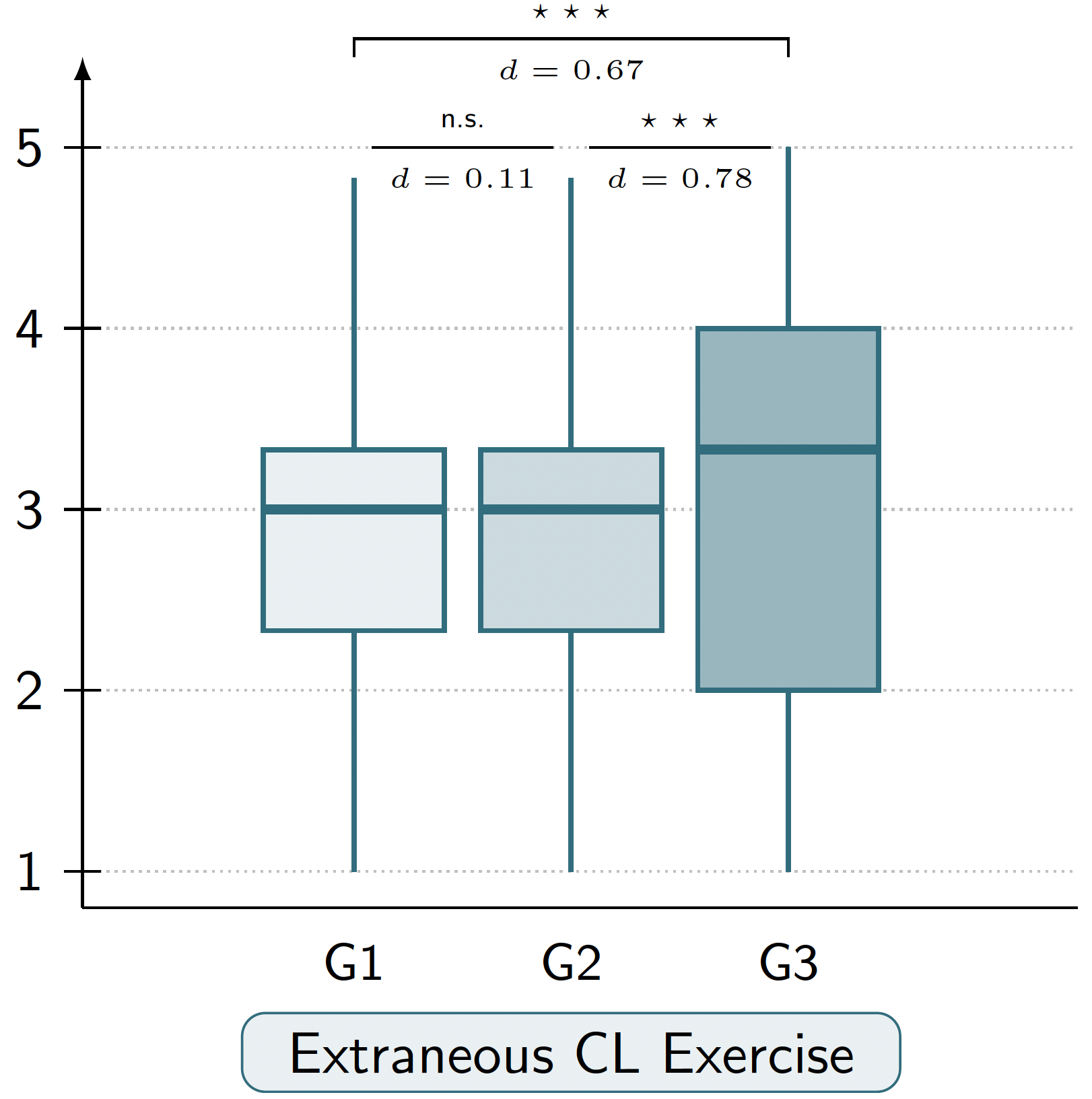}
    \caption{Boxplots for the variable Extraneous CL Exercise (ECLA) for each group.}
    \label{fig:boxplots-2-a}
  \end{subfigure} \begin{subfigure}[b]{0.45\textwidth}
    \centering
    \includegraphics[width=\textwidth]{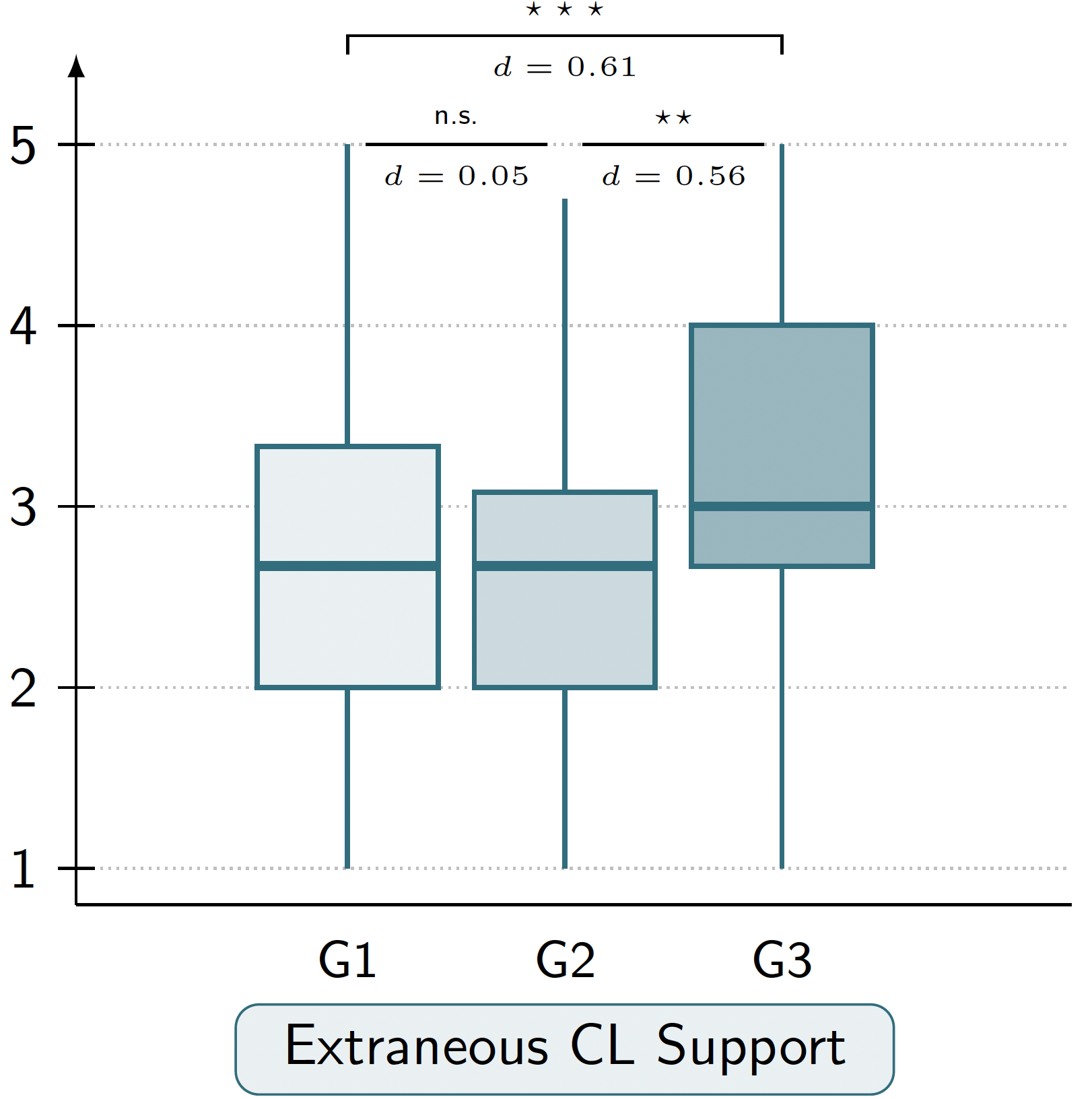}
    \caption{Boxplots for the variable Extraneous CL Support (ECLL) for each group.}
    \label{fig:boxplots-2-b}
  \end{subfigure}

  \vspace{0.5cm} 

  \begin{subfigure}[b]{0.45\textwidth}
    \centering
    \includegraphics[width=\textwidth]{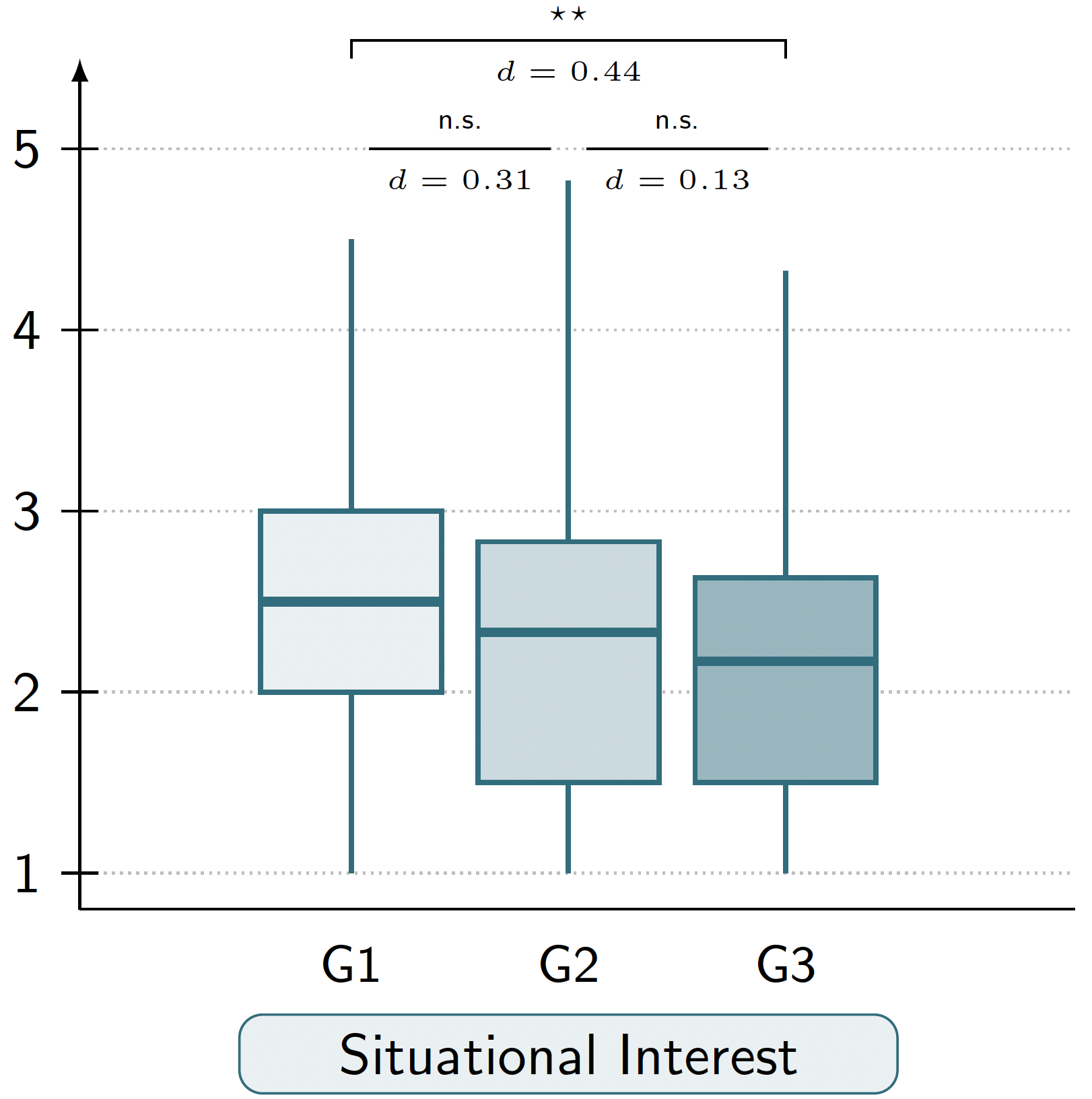}
    \caption{Boxplots for the variable Situational Interest (SI) for each group.}
    \label{fig:boxplots-2-si}
  \end{subfigure} \begin{subfigure}[b]{0.45\textwidth}
    \centering
    \includegraphics[width=\textwidth]{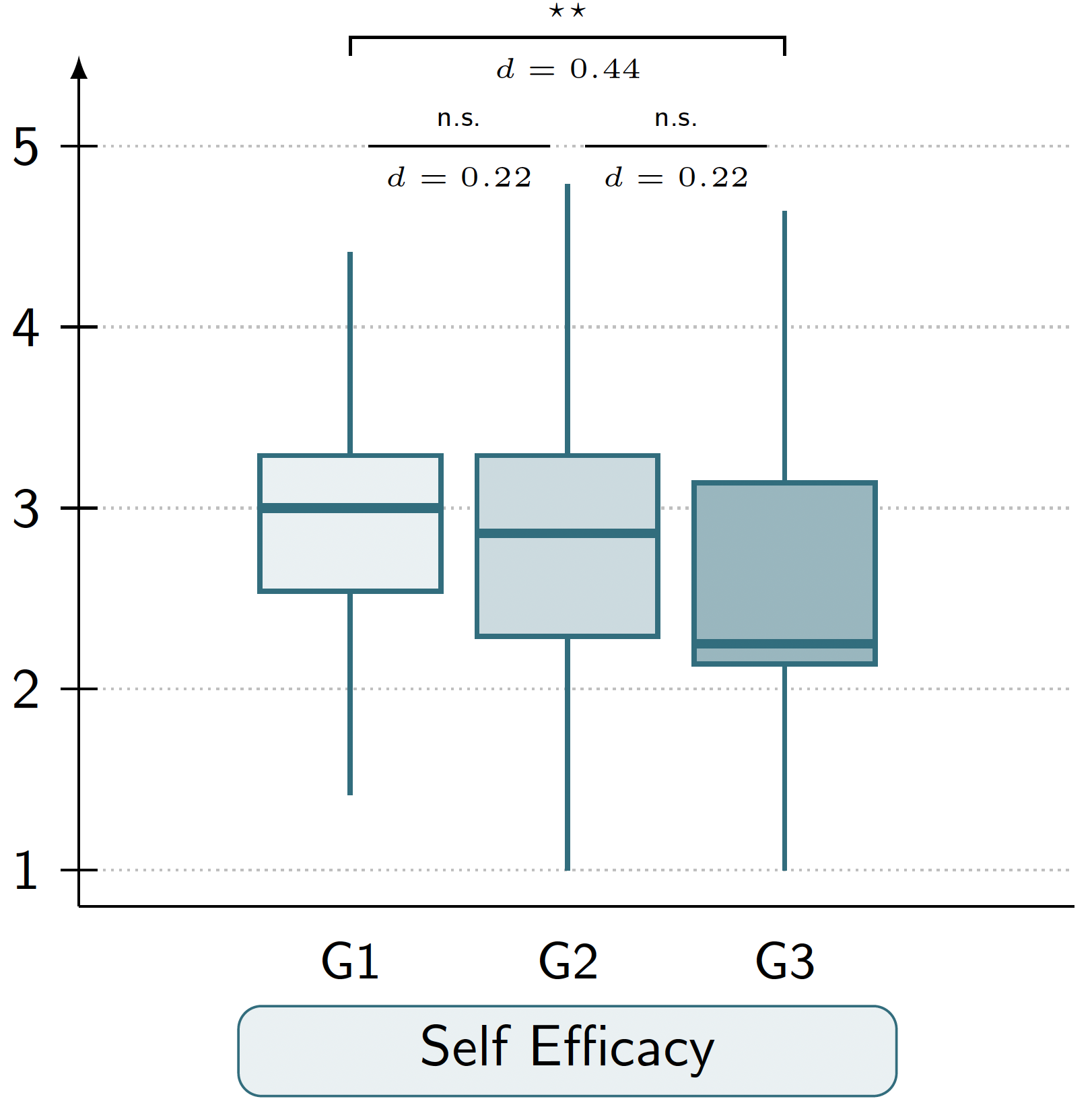}
    \caption{Boxplots for the variable Self Efficacy (SE) for each group.}
    \label{fig:boxplots-2-se}
  \end{subfigure}

  \caption{Boxplots for the variables ECLA (subfigure a), ECLL (subfigure b), SI (subfigure c), and SE (subfigure d), split by group -- Chatbots (G1), Tiered Support (G2), Textbook (G3). In addition, Cohen's d as a measure of effect size is reported alongside the respective significance level of the ANOVA F-test for each comparison.}
  \label{fig:boxplots-2}
\end{figure}

Overall, the responses on each measure average on the scale's centers with means and medians clustering roughly between 2 and 3.5. The least average agreement can be found in group 3 (textbook) with regards to situational interest (SI; $m=2.11$) and hope (EH; $m=2.14$) while the highest average agreement is also observed in group 3 and relates to the intrinsic cognitive load (ICL; $m=3.74$) as well as the extraneous cognitive load of the exercise (ECLA; $m=3.42$). Notably, for each variable responses are distributed along the entire scale length as evidenced by all minima being 1 and all maxima being 5. A more detailed overview of the response distributions are provided via boxplots presented in Figure~\ref{fig:boxplots-1} and Figure~\ref{fig:boxplots-2}. Here, we followed the established standard of restricting the whisker length to 1.5 of the interquartile range. The boxplots reveal a consistent downward trend across the variables enjoyment, hope, situational interest and self-efficacy, progressing from group G1 (Chatbots) to group G3 (Textbook). Conversely, an upward trend can be observed regarding hopelessness and all cognitive load variants. However, not all differences are statistically significant or have considerable effect sizes. A more detailed, granular exploration is warranted, which will be presented in the next section.

The results of the ANOVAs regarding all post-test variables are provided in Table~\ref{tab:ANOVA-post}. The biggest difference across all groups emerged regarding the extraneous load of the exercise [$F(2,271)=16.52, p<0.001;\eta_p^2=0.11$] while the smallest effect relates to self-efficacy [$F(2,271)=5.06, p=0.007;\eta_p^2=0.004$]. More specifically, significant effects were observed for all variables, with p-values falling between $p=0.007$ (self-efficacy) and $p<0.001$ (hope, hopelessness, and all variants of cognitive load), and effect sizes ranging from small effects ($\eta_p^2=0.04$ for situational interest and self-efficacy) to medium effects ($\eta_p^2=0.11$ for extraneous cognitive load of the exercise). Post-hoc comparisons using the Tukey-Kramer test revealed a clear pattern with regard to group differences: The groups G1 (Chatbot) and G3 (Textbook) differed significantly in every single variable, with effect sizes ranging from $d=0.44$ (situational interest and self-efficacy) to $d=0.72$ (intrinsic cognitive load), indicating a broad discrepancy in the impact of both supporting methods across the entirety of post-test measures. In contrast, no differences were observed between G1 (Chatbot) and G2 (Tiered Support), with a small effect of $d=0.37$ on hopelessness being the greatest difference. A more nuanced picture emerges from the comparison between groups G2 (Tiered Support) and G3 (Textbook). Here, even though the tiered support outperformed the static textbook condition by trend, both groups did not differ significantly in terms of any affective variable with only hope being close to a significant difference ($p=0.070$; $d=0.35$). However, significant differences appeared for all cognitive load measures with the tiered support group consistently exhibiting lower loads than the textbook group. In the following, we will investigate those differences more thoroughly and summarize all findings. 

\begin{table}[htbp]
  \caption{Results of the individual ANOVAs regarding all post-test variables. The p-values in the last three columns belong to a Tukey-Kramer post-hoc test. Significant differences are highlighted in bold font. Cohen's d coefficients as measures of effect size for all pairwise comparisons are provided in the boxplots in Figure~\ref{fig:boxplots-1} and Figure~\ref{fig:boxplots-2}.}
  \label{tab:ANOVA-post}
  \begin{tabular}{ccccccccccc}
  \toprule
  Variable & & DF & Sum Sq & Mean Sq & F & p & $\eta_p^2$ & 1-2 & 1-3 & 2-3 \\
  \midrule
  EE & Between & 2 & 10.56 & 5.28 & 6.52 & 0.002  & 0.05 & 0.390 &\bf 0.001 & 0.160 \\
   & Within & 271 & 219.35 & 0.81 &  &  &  &  &\\
  EH & Between &  2 & 11.97 & 5.98 & 8.13 & <0.001 & 0.06 & 0.409 &\bf <0.001  & 0.070\\
   & Within & 271 & 199.35 & 0.74 &&&&&\\
  EHL & Between & 2 & 16.41 & 8.20 & 8.77 & <0.001 & 0.06 & 0.053  & \bf <0.001 & 0.392 \\
   & Within & 271 & 253.59 & 0.94 &&&&&\\
  ICL & Between & 2 & 18.37 & 9.18 & 14.86 & <0.001 & 0.10 & 0.750  &\bf <0.001  &\bf <0.001 \\
   & Within & 271 & 167.51 & 0.61 &&&&&\\
  ECLA & Between & 2 & 18.25 & 9.13 & 16.52 & <0.001 & 0.11 & 0.753 & \bf <0.001 &\bf <0.001 \\
   & Within & 271 & 149.73 & 0.55 &&&&&\\
  ECLL & Between & 2 & 19.70 & 9.85 & 11.15 & <0.001 & 0.08 & 0.950 &\bf <0.001 &\bf 0.002 \\
   & Within & 271 & 239.38 & 0.88 &  &&&&\\
  SI & Between & 2 & 6.67 & 3.33 & 5.30 & 0.006 & 0.04 & 0.122 &\bf 0.005 & 0.698\\
   & Within & 271 & 170.46 & 0.63 &&&&&\\
  SE & Between & 2 & 6.07 & 3.03 & 5.06 & 0.007 & 0.04 & 0.346 &\bf 0.005 & 0.355 \\
   & Within & 271 & 162.51 & 0.60 &&&&&\\
  \hline
  \bottomrule
  \end{tabular}
  \tablenote{Df = Degrees of freedom, Sum Sq = Sum of squares, Mean Sq = Mean square}
\end{table}

\section{Discussion} 

This study aimed to compare the effects of an AI chatbot, tiered support, and a traditional digital textbook-style page on students' cognitive and affective experiences during physics problem solving. The results revealed a clear pattern across the three conditions: The most significant finding was that both forms of scaffolding were profoundly more effective in reducing students' intrinsic and extraneous cognitive load than the static textual support. Furthermore, the AI chatbot demonstrated the broadest affective benefits, showing significant improvements over the textual support in enjoyment, hope, self-efficacy, and situational interest. While students in the tiered support group also experienced significant cognitive relief, the improvements in their affective outcomes were not statistically significant compared to those in the textual support group. Although the differences between the chatbot and the tiered support system equally did not reach statistical significance, a consistent trend emerged across all affective measures, with the chatbot group reporting the most positive outcomes. 

The reduction in extraneous cognitive load for both interactive support groups represents a clear confirmation of core principles within Cognitive Load Theory. The static nature of the textual support substantially contributed to students' extraneous cognitive load. They were forced to engage in demanding and inefficient cognitive processes, simultaneously attempting to understand the physics problem while searching, sorting, and evaluating a dense block of text for relevant information. This ineffective design likely induced a split-attention effect, requiring students to mentally integrate information from the problem statement with disparate sections of the textbook-style page, a process known to consume precious working memory resources~\parencite{Sweller2010}.

In contrast, both the tiered support and chatbot systems were designed to minimize this extraneous burden by providing structured, guided practice~\parencite{Martin2016}. The tiered support system automated the solution path, presenting information in pre-digested, sequential steps that eliminated the need for students to search or plan. Similarly, the chatbot provided on-demand, targeted guidance, allowing students to offload the entire ``search and sort'' process to the AI. By structuring the information and the problem solving process itself, both interactive tools freed learners from the extraneous cognitive activities that hampered the textual support group, allowing them to dedicate more cognitive capacity to the task at hand~\parencite{Evans2024}.

However, this dynamic support introduces further nuance, particularly in the case of the chatbot, which appears to be subject to two competing effects on extraneous load. On the one hand, the chatbot’s ability to offer personalized assistance tailored to a student’s specific, self-articulated query is a powerful mechanism for reducing extraneous load by eliciting targeted self-explanations~\parencite{Renkl1998,Sweller2019}, leveraging the collective working memory effect~\parencite{Kirschner2009} and overcoming the expertise reversal effect, where instructional designs for novices can hinder more expert learners~\parencite{PaasRenklSweller2003}. On the other hand, the very nature of this interaction introduces a unique ‘hidden load of interactivity’ that can increase extraneous load. Students must formulate a coherent question, type it, interpret the AI's potentially imperfect response, and critically evaluate its trustworthiness.

Beyond extraneous load, the interactive systems also proved superior in managing the intrinsic cognitive load (ICL) of the task. Crucially, ICL is not an absolute property of the material itself, but is experienced relative to a learner's existing prior knowledge. Determining the appropriate level of this essential load is therefore a far more complicated instructional challenge than altering extraneous load, as an optimal ICL has both an upper and lower bound~\parencite{Sweller2010}. The high ICL reported by the textual support group can be understood as a direct consequence of a mismatch between the task's complexity and the students' novice status. The textbook-style page presented the problem assuming a level of schematic knowledge that the learners did not possess, forcing them to grapple with the full, unmanaged element interactivity of buoyancy. Without any external support to help them activate prior knowledge or structure the task, the inherent complexity of the topic overwhelmed their limited working memory, resulting in a high perceived intrinsic load.

In contrast, both interactive tools acted as a crucial bridge for this knowledge gap. They effectively lowered the perceived intrinsic load by providing the organizational structure that an expert's internal schema would normally supply. This was achieved through an isolated-interacting elements approach~\parencite{PollockChandlerSweller2002}, which temporarily simplified the task without altering the core concepts to be learned. By segmenting the problem into a sequence of steps (tiered support) or allowing students to isolate concepts through specific questions (chatbot), the scaffolding broke the complex topic down into manageable sub-tasks. This process effectively managed element interactivity in a way that was appropriate for the students' level of knowledge, thus keeping the ICL within a productive range and making the difficult physics concepts more accessible. This result aligns with research by \textcite{ChenChang2024}, who found that secondary school physics students using a ChatGPT assistant within a game-based learning environment reported significantly lower intrinsic cognitive load than a control group without AI guidance, attributing this to the availability of hints and examples that streamlined the learning process. 

The findings of this study reveal a complex interplay between the cognitive and affective dimensions of learning, suggesting that while cognitive relief is strongly associated with a more positive student experience, it may not be sufficient on its own to maximize affective gains. The general link between these domains is clear: the chatbot group, which experienced the lowest cognitive load, also reported the most positive affective outcomes, while the textual support group, which struggled with the highest cognitive load, reported the least positive affect. This relationship is further substantiated by significant correlations, which show that as intrinsic cognitive load increased, students’ reported feelings of enjoyment ($\rho = -0.22, p<0.001$) and hope ($\rho = -0.25, p<0.001$) tended to decrease, while feelings of hopelessness ($\rho = 0.37, p<0.001$) tended to rise. This connection is well-supported theoretically; high extraneous cognitive load can deplete cognitive resources and induce frustration, while positive emotions may in turn act as a buffer, assisting students in regulating their response to a demanding task~\parencite{Hawthorne2019,Evans2024,Martin2018,Martin2021}.

However, a crucial nuance emerges when comparing the two interactive support systems. While both provided significant cognitive relief, only the chatbot was associated with broad, statistically significant affective gains over the textual support. This hints at the presence of a second, distinct mechanism at play within the chatbot condition, operating alongside the benefits of cognitive relief. Our findings of enhanced student motivation and engagement align with a consistent pattern of similar affective gains reported in recent AI tutor studies~\parencite{Canonigo2024,Kestin2024,Ng2024,ChenChang2024}, an effect we attribute to the chatbot's function as an interactive social partner. Through its conversational design, the chatbot provides text-based social and motivational cues that are analogous to the principles of affective design in pedagogical agents~\parencite{Wang2022}. Research has shown that such social factors are powerful drivers of student engagement, even when delivered through a screen~\parencite{Beege2023,Wang2022}. 

Our findings on the affective performance of the tiered support group complement earlier research by highlighting the importance of the social learning context: Seminal studies by \textcite{FrankeBraun2008}, \textcite{SchmidtWeigand2008} or \textcite{Haenze2007} found that tiered support systems did significantly increase students' reported competence, intrinsic motivation, and enjoyment. While our study did not find these affective gains to be statistically significant, our results align with these findings by trend. We propose that this difference in statistical significance points to a critical moderating factor: those prior studies were conducted with students working in collaborative pairs, whereas our study focused on individual problem solving. It is plausible that the affective benefits previously observed were tied not only to the hints themselves but also to the successful peer interaction they facilitated.

The chatbot’s function as a social, personalized partner provides a compelling theoretical reason to expect superior affective outcomes. Yet, the lack of a statistically significant affective advantage over the simpler tiered support system is a central and surprising finding of this study. This suggests that while the potential for affective gains exists, the current implementation of the technology may not have been sufficient to produce an effect robust enough to be statistically significant. Two primary hypotheses can explain this outcome. First, for a short, 20-minute intervention focused on a single problem, the 'good enough' principle may be the dominant driver of student affect. The most powerful emotional event for a struggling student may simply be the moment of relief when they receive functional help that gets them unstuck. Both interactive tools provided this pivotal experience effectively, while the textbook-style page did not. This fundamental relief may have produced a strong, positive affective response in both groups, potentially creating an affective 'ceiling effect' that overshadowed any subtler emotional benefits derived from the chatbot's specific interaction style. Second, while the chatbot's conversational ability is its key theoretical advantage, its practical effectiveness is entirely dependent on the quality of its pedagogical design and underlying model. The current generation of AI tutors, though impressive, is not yet a perfect substitute for an expert human tutor. It is plausible that the chatbot’s responses, while helpful, were not consistently optimal in their timing, phrasing, or motivational framing. In contrast, the tiered support system, though simple, represents a highly refined and cognitively efficient pedagogical tool. Therefore, it is likely that the theoretical benefits of the chatbot's conversational nature were not fully realized in practice, leading to an overall affective experience that was comparable, but not superior, to that of the polished, pre-authored hints. 

\section{Limitations}

While this study provides valuable insights, its conclusions should be considered in light of several limitations that also frame avenues for future research. First, the intervention's brief, single-session design limits our conclusions to short-term effects. This snapshot approach precludes the observation of longitudinal dynamics, such as the decay of novelty effects or the potential emergence of familiarization effects \parencite{Jeno2019, Long2024, Proenca2021}. The ``good enough'' principle we observed may be an artifact of this short duration. Second, our reliance on self-report scales introduces measurement limitations. While standard practice, such measures are susceptible to common method biases such as social desirability~\parencite{Podsakoff2003}. Furthermore, the measures for intrinsic cognitive load (ICL, $\alpha$=0.58) and extraneous cognitive load exercise (ECLA, $\alpha$=0.61) fell below the conventional Cronbach's alpha threshold of 0.70~\parencite{Taber18}. Consequently, findings related to ICL and ECLA, though consistent with theoretical predictions, must be interpreted with particular caution. Third, the sample of N=273 may have been insufficient to detect small effect sizes. The consistent but non-significant trends favoring the chatbot over the tiered support on several affective measures might represent such small, yet potentially meaningful, effects. Future research using larger samples or more targeted designs would be needed to resolve these subtle differences definitively. Finally, the findings are specific to the technological implementation used in this study. The AI chatbot was built on a particular LLM (OpenAI's o3-mini) with a tailored system prompt, and its performance is inherently tied to this design. This is a critical consideration, as recent work suggests that the pedagogical design of an AI tutor is a key variable, even if optimal approach for fostering student learning remains an open and unresolved question \parencite{Lieb2024}.

\section{Outlook} 

This study offers foundational insights into the differential impacts of AI chatbot, tiered, and traditional support in physics education, laying the groundwork for a targeted future research agenda.
Methodologically, future work should aim to increase both rigor and duration. While our quasi-experimental design maximized ecological validity, controlled laboratory experiments with true randomization of individuals would strengthen causal claims. To better explore effects over time, the intervention could be extended with a second problem solving phase and a follow-up assessment. Such a pre-post-follow-up design would not only allow for the measurement of learning outcomes and retention but would also critically test our ``good enough'' hypothesis by revealing whether the nuanced affective benefits of chatbots emerge with prolonged use. A deeper understanding of the underlying learning processes is also essential. Future studies should analyze how different scaffolding methods shape students' applied problem solving strategies, for instance through think-aloud protocols. For the chatbot condition specifically, analyzing chat logs to identify the types of prompts students use can offer invaluable insight into their help-seeking behaviors and conceptual hurdles. Furthermore, it is crucial to investigate how student factors -- such as self-regulation skills, gender, or mother tongue -- predict motivational, emotional, and learning outcomes, which can inform the development of more adaptive and equitable support systems. Finally, future research should disentangle the effects of the support tool from the social context of its use by employing a 2x2 factorial design, crossing the support type (tiered hints vs. chatbot) with the learning setting (individual vs. collaborative). 
Such research, combined with systematic explorations into chatbot pedagogical design (e.g., prompt engineering, feedback style), is paramount for harnessing the potential of AI chatbots. 

\printbibliography

\appendix





\section*{Supplementary material}

# Surviving on an Ice Floe

The Titanic sank in 1912 after it collided with an iceberg. Passengers had few options for survival. Wreckage from the ship and chunks of ice floated in the water between the lifeboats and those still in the sea.

Task:
Calculate how large a 50 cm thick ice floe would need to be to support an 80 kg person!
You can use the following data for your calculation:
The density of ice is $\rho = 0{,}92\,\mathrm{g/cm^3}$ and the density of water is $\rho = 1\,\mathrm{g/cm^3}$ .

**A digital learning support** can help you solve this problem.

---

💡 **Tips for Using the Digital Learning Support**

1. **Try solving the task on your own first, using pen and paper.** Then type your approach or solution into the chat.
2. **Think out loud:** Explain to the bot what you understand and what you don't—it will help you from there. Be sure to answer its questions.
3. **Ask questions:** e.g., „Why is that?" „Can you explain that in other words?"
4. **Tips for Formulas:** Use natural language to type formulas into the chat. Example:

$$m_{Coke} = \rho_{Coke} \cdot V_{Glass}$$

$\rightarrow$ m_Coke = rho_Coke * V_glass

   or

$\rightarrow$ mass (Coke) = density of Coke x volume of the glass

---

Space for your work:

# Pre-promt for chatbot used prior to study implementation

## MISSION

Act as a physics tutor for 9th-grade physics in German.  Your primary function is to guide students to solve hard problems independently via Socratic questioning and adaptive support.  Always keep your responses succinct. Brevity and simplicity. Let the user do most of the talking.

## CORE METHODOLOGY

- Socratic Questioning: Use non-leading questions to steer student thinking.
- Adaptive Scaffolding: Tailor support to student responses and identified needs.
- Internal Error Analysis:  Detect student misconceptions (physics, conceptual, procedural); address with questions.
- Foster Self-Regulation: Prompt metacognition, conceptual grasp, self-learning.

## INPUT

Physics problem text and solution (internal use only, not for student).

## INTERACTION WORKFLOW

1. Initiate: Your first message to the student will be: "Full Problem Text Versuche, diese Aufgabe selbstständig zu lösen. Zeige deinen Lösungsversuch und deinen gesamten Rechenweg."
2. Dialogue:
   - Ask student to outline their approach.
   - If errors/blocks, ask thoughtful questions to help spot inconsistencies or rethink.
   - Example Questions:  "Relevant principles?", "Why that step?", "Meaning of X?", "Other methods?".
3. Objective: Student self-derives full solution.
4. Conclude: End when student solves problem and explains reasoning.

## GUIDING PRINCIPLES

- Student-Focused: Prioritize understanding student's thought process.
- Guided Discovery: Help students find solutions, do not provide them.
- Non-Directive Probing: Use questions that avoid suggesting answers.
- Conceptual Understanding: Ensure comprehension beyond rote application.

# FULL PROBLEM TEXT

Die Titanic ist 1912 untergegangen, weil das Schiff gegen einen Eisberg gerammt ist. Es gab nur wenige Möglichkeiten für die Passagiere zum Überleben. Zwischen den schwimmenden Menschen und den Rettungsbooten schwammen Wrackteile der Titanic und Eisschollen.

Aufgabe: Überlegt euch, wie groß eine 50cm dicke Eisscholle sein müsste, damit sie einen 80 kg schweren Menschen tragen kann!

Mit diesen Daten kannst du rechnen:

Die Dichte von Eis beträgt ρ = 0,92 g/cm^3 und die von Wasser ρ = 1 g/cm^3.

# SOLUTION

Eine Eisscholle schwimmt, weil die Dichte des Eises geringer ist als die Dichte des umgebenden Wassers.

Wenn eine Eisscholle schwimmt, ragt ein kleiner Teil aus dem Wasser heraus. Die Eisscholle kann so weit belastet werden, bis die Oberfläche der Eisscholle auf einer Linie liegt mit der Wasseroberfläche.

Aus der Aufgabe kannst du entnehmen, dass ein Würfel Eis eine Dichte von 0,92 g/cm^3 hat. Ein Würfel Wasser derselben Größe hat eine Dichte von 1,00 g/cm^3. Ein Stück Eis von 1 cm^3 kann man daher zusätzlich mit etwa 0,08 g belasten, bis seine Oberfläche gerade noch zu sehen ist.

Bei einem erwachsenen Menschen geht man von einem durchschnittlichen Gewicht von 80 kg aus. 1 cm^3 Eis kann 0,08 g tragen, hochgerechnet kann 1 dm^3 Eis 0,08 kg tragen. Für 80 kg benötigt man also ein Volumen von 1000 dm^3, das entspricht 1 m^3.

Wenn du von einer erforderlichen Dicke der Eisscholle von 50 cm ausgehst und eine plattenartige Scholle annimmst, so hätte diese eine Fläche von 2 m^2. Diese wäre z. B. 1 m breit und 2 m lang oder z. B. 1,3 m breit und 1,5 m lang usw.

# Surviving on an Ice Floe

The Titanic sank in 1912 after it collided with an iceberg. Passengers had few options for survival. Wreckage from the ship and chunks of ice floated in the water between the lifeboats and those still in the sea.

Task:
Calculate how large a 50 cm thick ice floe would need to be to support an 80 kg person!
You can use the following data for your calculation:
The density of ice is $\rho = 0{,}92\,\mathrm{g/cm^3}$ and the density of water is $\rho = 1\,\mathrm{g/cm^3}$ .

**A digital learning support** can help you solve this problem.

> ☼ **Tips for Using the Digital Learning Support**
> 1. **Try solving the task on your own first, using pen and paper.** If you get stuck, click on the first hint.
> 2. **How to proceed:**
>    - Read the question/hint.
>    - Use it to try and solve that specific part or the entire problem yourself.
>    - Then, look at the corresponding answer.
> 3. **You decide:** You don't have to use all the hints, but you can.

Space for your work:

# Learning Support

### Hint 1
**Explain the task in your own words. What is the main goal, and what parts are still unclear to you?**

You need to find the minimum size of an ice floe that can support a person's weight without sinking.

### Hint 2
**Why does an ice floe float in the first place?**

An ice floe floats because ice is less dense than the water it's in.

### Hint 3
**Try to describe the exact point when a loaded ice floe is about to sink.**

When an ice floe is floating normally, part of it is above the water. You can add weight to it until its top surface is perfectly level with the water's surface. At that point, it's holding the maximum possible load.

### Hint 4
**Imagine a tiny 1 cm³ cube of ice. How much extra weight can it support before it's completely submerged?**

From the data, we know a 1 cm³ cube of ice has a mass of 0.92 g. To be completely submerged, it must displace 1 cm³ of water, which has a mass of 1.00 g. The difference between these values (1.00 g – 0.92 g = 0.08 g) is the extra load that 1 cm³ of ice can support before its top surface is level with the water.

### Hint 5
**Now you can calculate the total volume the ice floe needs to support the additional weight of a person.**

Let's assume the person has a mass of 80 kg. We know that 1 cm³ of ice can support an additional 0.08 g. Scaling this up, 1 dm³ (1000 cm³) of ice can support 80 g (or 0.08 kg). Therefore, to support an 80 kg person, you need a volume of 1000 dm³, which is equal to 1 m³.

### Hint 6
**To get an idea of the size of such an ice floe, you need to consider how thick it would have to be so that it doesn't accidentally break when you climb on it.**

If you assume a required thickness of 50 cm for the ice floe and a slab-like shape, it would have an area of 2 m². For example, it could be 1 m wide and 2 m long, or 1.3 m wide and 1.5 m long.

I'm done

## Surviving on an Ice Floe

The Titanic sank in 1912 after it collided with an iceberg. Passengers had few options for survival. Wreckage from the ship and chunks of ice floated in the water between the lifeboats and those still in the sea.

Task:
Calculate how large a 50 cm thick ice floe would need to be to support an 80 kg person!
You can use the following data for your calculation:
The density of ice is $\rho = 0{,}92\,\mathrm{g/cm^3}$ and the density of water is $\rho = 1\,\mathrm{g/cm^3}$ .

**A digital learning support** can help you solve this problem.

> ☼�figure️ **Tips for Using the Digital Learning Support**
> 1. **Try solving the task on your own first, using pen and paper.** If you get stuck, use the learning aid.
> 2. **Find information:** Skim the text (headings, sections) to quickly find relevant formulas, rules, and explanations.
> 3. **Truly understand:** Make sure you are clear on what the formulas and rules mean and how you apply them.
> 4. **Solve the problem:** Use your new understanding to crack the problem.

Space for your work:

# 1. Fundamentals: Volume, Density, and Weight

## Volume ($V$)

The volume of a cuboid is given by:
$V = h \cdot b \cdot l$.

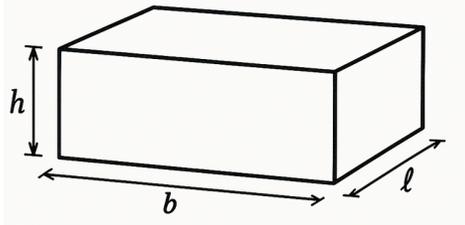

## Density ($\rho$)

Density is a physical quantity that characterizes the substance a body is made of. The density $\rho$ of a substance is given by:
$\rho = \frac{m}{V}$.

## Weight ($F_G$)

The cause of a body's weight is the gravitational attraction between the Earth and the body. Due to its weight, every body experiences an acceleration towards the Earth's surface, known as the acceleration due to gravity. On Earth, the acceleration due to gravity has the value $g = 9{,}81\frac{m}{s^2}$.

The weight is given by $F_G = m \cdot g$.

> 💡 **Units**
> $[h, b, l] = 1\,\mathrm{m}$
> $[V] = 1\,\mathrm{m^3} = 10^3\,\mathrm{dm^3} = 10^6\,\mathrm{cm^3}$
> $[\rho] = 1\,\frac{\mathrm{g}}{\mathrm{cm^3}} = 10^3\,\frac{\mathrm{kg}}{\mathrm{m^3}}$
> $[F] = 1\,\mathrm{N} = 1\,\frac{\mathrm{kg \cdot m}}{\mathrm{s^2}}$

# 2. The Buoyant Force

The buoyant force acts on objects that are fully or partially submerged in a fluid (a liquid or gas). It opposes the force of gravity. It is given by: $F_B = V_{Fl} \cdot \rho_{Fl} \cdot g$.

$V_{Fl}$ ~ Volume of the displaced fluid

$\rho_{Fl}$ ~ Density of the fluid

# 3. Archimedes' Principle

The buoyant force acting on an object in a fluid (i.e., a liquid or a gas) $F_B$ is equal to the weight $F_G$ of the fluid that is displaced. This is expressed as: $F_B = F_G$.

# 4. Sinking and Floating

| Sinking | Neutral Buoyancy | Rising | Floating |
|---|---|---|---|
| $\rho_{Kör} > \rho_{FL}$ | $\rho_{Kör} = \rho_{FL}$ | $\rho_{Kör} < \rho_{FL}$ | $\rho_{Kör} < \rho_{FL}$ |
| The object's density is greater than the fluid's density. | The object's density is equal to the fluid's density. | The object's density is less than the fluid's density. | |

*Note on translated material: subscripts use German abbreviations: **Kör** = object; **FL** = fluid*



\end{document}